\pdfoutput=1 
\documentclass[a4paper, 11pt]{article}
\PassOptionsToPackage{table}{xcolor}

\usepackage{jheppub} 

\usepackage[T1]{fontenc} 
\usepackage{tangocolors}
\usepackage{enumitem}
\usepackage{tikz}
\usetikzlibrary{arrows, decorations,backgrounds, patterns}
\usetikzlibrary{decorations.pathreplacing ,decorations.markings}
\usepackage{amssymb}
\usetikzlibrary{decorations.pathmorphing,backgrounds,shapes,arrows,shadows}
\usetikzlibrary{patterns.meta}
\usetikzlibrary{patterns,decorations.pathmorphing}
\tikzset{
    snake it/.style={decorate, decoration=snake}
}
\usepackage{pgfplots}
\pgfplotsset{compat=1.11}
\usepgfplotslibrary{fillbetween}
\usetikzlibrary{intersections}
\pgfdeclarelayer{bg}
\pgfsetlayers{bg,main}
\tikzset{zigzag/.style={decorate,decoration=zigzag}}
\tikzset{snake it/.style={decorate, decoration=snake}}
\makeatletter
\def\@hex@@Hex#1%
 {\if a#1A\else \if b#1B\else \if c#1C\else \if d#1D\else
  \if e#1E\else \if f#1F\else #1\fi\fi\fi\fi\fi\fi \@hex@Hex}
\makeatother

\usepackage[all]{xy}

\makeatletter  
\gdef\@fpheader{}  
\makeatother 
\usepackage[percent]{overpic}
\usepackage{slashed}
\usepackage{wrapfig}
\usepackage{tabu}
\usepackage{diagbox}
\usepackage{mathrsfs,amsmath,amssymb,amsthm,amsfonts,tikz,graphicx,accents,hyperref, color}
\usepackage{dsfont,epiolmec, latexsym, stmaryrd, comment}
\usepackage{slashed,ccaption}
\usepackage{mathrsfs, calligra}
\usepackage{leftidx}
\usepackage{import}
\usepackage{multirow}
\usepackage{amsfonts}
\usepackage{pifont}
\usepackage{tabularx}
\usepackage{cancel}
\usepackage[utf8]{inputenc}
\usetikzlibrary{intersections,calc}
\usepackage{ifthen}
\usepackage{amsmath}
\usepackage{cancel}
\usepackage{caption} 
\usepackage{subcaption} 
\DeclareMathOperator{\extdm}{d}
\newcommand{\extd}{\extdm \!}

\usetikzlibrary{patterns.meta}
\usepackage{array}
\hypersetup{ linktoc=all,
    colorlinks, linkcolor={palatinateblue},
    citecolor={brightpink}, urlcolor={amaranth}
}

\graphicspath{{Images/}}

\def\sideremark#1{\ifvmode\leavevmode\fi\vadjust{\vbox to0pt{\vss
 \hbox to 0pt{\hskip\hsize\hskip1em
 \vbox{\hsize2cm\tiny\raggedright\pretolerance10000
 \noindent #1\hfill}\hss}\vbox to8pt{\vfil}\vss}}}%
\DeclareSymbolFont{extraup}{U}{zavm}{m}{n}
\DeclareMathSymbol{\varheart}{\mathalpha}{extraup}{86}
\DeclareMathSymbol{\vardiamond}{\mathalpha}{extraup}{87}
\makeatletter
\renewcommand*{\@fnsymbol}[1]{\ensuremath{\ifcase#1\or \clubsuit \or \vardiamond \or \varheart\or
    \spadesuit\or \mathparagraph\or \|\or **\or \dagger\dagger
    \or \ddagger\ddagger \else\@ctrerr\fi}}
\makeatother

\definecolor{rosy}{RGB}{230,235,252}
\definecolor{myframetitle}{RGB}{90,89,170}
\definecolor{myblocktitle}{RGB}{140,185,249}
\definecolor{mytitle}{RGB}{10,80,26}

\definecolor{darkgreen}{RGB}{27,130,45}
\definecolor{darkblue}{rgb}{0,0,0.3}
\definecolor{darkred}{rgb}{0.7,0,0}

\definecolor{light gray}{RGB}{220,220,220}
\definecolor{dark purple}{RGB}{108,0,217}
\definecolor{pink}{RGB}{190,20,100}
\definecolor{orang}{RGB}{193,63,0}
\definecolor{green}{RGB}{11,98,17}
\definecolor{darkpink}{RGB}{153,0,76}
\definecolor{bluegreen}{RGB}{0,102,102}
\definecolor{greenlagan}{RGB}{0,102,0}
\definecolor{redgreen}{RGB}{102,102,0}
\definecolor{Redgreen}{RGB}{153,76,0}
\definecolor{vividviolet}{rgb}{0.62, 0.0, 1.0}
\definecolor{amaranth}{rgb}{0.9, 0.17, 0.31}
\definecolor{palatinateblue}{rgb}{0.15, 0.23, 0.89}
\definecolor{brightpink}{rgb}{1.0, 0.0, 0.5}
\definecolor{cornflowerblue}{rgb}{0.39, 0.58, 0.93}
\definecolor{deepcarminepink}{rgb}{0.94, 0.19, 0.22}
\definecolor{radicalred}{rgb}{1.0, 0.21, 0.37}
\usepackage[most]{tcolorbox}

\tcbset{highlight math style={left=02mm,right=02mm,top=-1mm,bottom=-1mm}} 
\usepackage{empheq}

\DeclareFontFamily{OT1}{rsfs}{}

\DeclareFontShape{OT1}{rsfs}{m}{n}{ <-7> rsfs5 <7-10> rsfs7 <10->rsfs10}{} 

\DeclareMathAlphabet{\mycal}{OT1}{rsfs}{m}{n}
\makeatletter \@addtoreset{equation}{section}

\begin{document}


\newcommand{\mytitle}{{\textbf{\centerline{\LARGE{Fracton and Non-Lorentzian Particle Duality:}}\\ \vskip 2mm \centerline{\Large{Gauge Field Couplings and Geometric Implications}}
}}}

\title{{\mytitle}}
\author[a]{M.M.~Ahmadi-Jahmani}
\author[b]{, A.~Parvizi }
\affiliation{$^a$ Department of Physics, Faculty of Science, Ferdowsi University of Mashhad,\\ P.O.Box 1436, Mashhad, Iran}
\affiliation{$^b$ University of Wroclaw, Faculty of Physics and Astronomy, Institute of Theoretical Physics,\\ Maksa Borna 9, PL-50-204 Wroclaw, Poland}

\emailAdd{mohammadmehdi.ahmadijahmani@mail.um.ac.ir,
aliasghar.parvizi@uwr.edu.pl,
}

\abstract{Fractons, characterized by restricted mobility and governed by higher-moment conservation laws, represent a novel phase of matter with deep connections to tensor gauge theories and emergent gravity. This work systematically explores the duality between fractons and non-Lorentzian particles—Carroll and Galilean—within electromagnetic (EM) fields. By constructing canonical actions for fractons in rank-2 gauge fields, we derive their equations of motion and demonstrate a new set of dualities between fractons and non-Lorentzian particles in gauge fields. The algebraic underpinnings of these dualities are clarified through symmetry analyses, revealing structural parallels between the fracton and Carroll/Galilean algebras. Furthermore, by gauging the fracton algebra, we develop a framework for coupling fracton gauge fields and background geometry, linking them to non-Lorentzian spacetimes and deriving the corresponding constraint on geometry. These results unify fracton dynamics with non-relativistic and ultra-relativistic limits of physics, offering insights into emergent gravity and exotic condensed matter systems.}    

\maketitle

\section{Introduction}
Fracton particles, also known as the fractonic phase of matter, represent a recent advancement in high-energy and condensed matter physics \cite{Pretko:2020cko,Nandkishore:2018sel,Gromov:2022cxa}. Fractons exhibit restricted mobility; some remain immobile when isolated. In other words, they cannot move without creating additional excitations \cite{Pretko:2016kxt,Pretko:2016lgv,Pretko:2017apy,Pretko:2017fbf,Pretko:2018jbi}, while others may form mobile bound states. These mobile bound states often have mobility only in specific directions, leading to terms such as ‘lineon’ and ‘planon \cite{Nandkishore:2018sel,Vadali:2023raq,Spieler:2023wkz,Zhang:2023eaw}.

Consequently, two primary approaches have been employed in describing Fractons. The first approach, initially introduced in Chamon’s seminal paper \cite{Chamon:2004lew} and further explored in other works \cite{Bravyi:2010jfq,Haah:2011drr,Bravyi:2013ort,Vijay:2016phm,Muhlhauser:2019rjg,Villari:2023myb}, leverages tools from quantum information theory to study exactly solvable spin and Majorana models \cite{Villari:2023myb,Prakash:2023ymm}.
The second approach, which is currently widely adopted, centers on investigating specific symmetric tensor gauge theories. This pioneering approach, introduced by Pretko \cite{Pretko:2016kxt}, establishes connections with distinct fields of physics, including elasticity theory and gravity \cite{Pretko:2017kvd,Pretko:2017fbf}.
These quasi-particles naturally couple with spatial two-index symmetric tensor gauge fields, denoted as $A_{ij}$. This implies a connection between dipole conservation and spin-2 fields, potentially even gravity \cite{Pretko:2016kxt,Pretko:2016lgv}. Researchers have investigated the internal dynamics of Fracton systems, exploring their potential links to emergent gravitational phenomena \cite{Afxonidis:2023pdq,Pretko:2017fbf}. The initial clue lies in the fact that gravity is also described by a symmetric tensor represented by the metric $g_{\mu\nu}$ \cite{Pretko:2017fbf}.

However, non-Lorentzian physics has recently gained interest \cite{Banerjee:2024jub}. Non-relativistic physics, which stems from the non-relativistic limit of relativistic physics (known as Galilean relativistic theory), was first introduced by Le Bellac and Levy-Leblond \cite{levy1965nouvelle,le1973galilean}. It finds applications in modern physics \cite{Taylor:2008tg,Nishida:2007pj,Son:2008ye,Goldberger:2008vg}. Another intriguing limit of relativistic symmetry is the Carroll symmetry, which arises when the speed of light is significantly smaller than that of other speeds. Levy-Leblond proposed this limiting case \cite{levy1965nouvelle}. Using group contraction from the Poincaré group (describing relativistic symmetry), we can derive both the Galilean and Carroll groups \cite{Duval:2014uoa,Bergshoeff:2022eog,Figueroa-OFarrill:2022nui,Figueroa-OFarrill:2022pus}. In particular, the Carroll group and the Galilean group are isomorphic in $d=2$ (where $d$ represents spatial dimension) \cite{Figueroa-OFarrill:2022nui}.

The connection between Fractons and non-Lorentzian particles has already been established. Spatially, in \cite{Figueroa-OFarrill:2023vbj}, the authors showed the duality between Carroll particles and Fractons. In \cite{Figueroa-OFarrill:2024ocf}, they described  Galilean particles and briefly mentioned the fact that massless Galilean particles can be connected to Fracton excitations which are known as Planons.

In this paper, our motivation is to explore deeper into this connection, specifically focusing on the electromagnetic couplings of Fractons and investigating the conditions affecting their mobility and finding their non-Lorentzian dual particle theory. We start by constructing and developing the theory of fractons in a gauge field, then employ a systematic approach (limiting procedure) to derive the non-Lorentzian particle dynamics and explore their dual description on the fractonic side. Explicitly, we demonstrate that static fractons in gauge fields exhibit duality with the electric sector of charged Carroll particles in electromagnetic (EM) fields, whereas mobile bound states of fractons (mobile dipoles) are dual to the electric and magnetic sectors of charged Galilean massless particles in the EM field. Additionally, we establish that the magnetic sector of Carroll particles is also dual to a certain model of mobile fracton dipoles. This intriguing duality between Fractons and non-Lorentzian particles prompts us to explore Fracton geometry alongside non-Lorentzian geometry, enhancing our understanding of their interconnectedness in the last section.  

\paragraph{Organization:} This paper opens with an introduction that provides a background on fracton particles and non-Lorentzian theories, setting the foundation for the exploration that follows. In Section~\ref{Mathematical description of fractons}, we develop the theory of fracton particles coupled to a rank-2 gauge field, enforcing invariance under fracton symmetry. Section~\ref{Non-Relativistic point particle} analyzes the standard Lorentzian point particle theory and its limiting transitions toward Carroll and Galilean regimes. This discussion is extended in Section~\ref{Non-Relativistic point particle in EM}, where we investigate particles interacting with gauge fields and identify potential non-Lorentzian behaviors at two distinct extreme limits, then the fracton/non-Lorentzian particle
duality is extended in Section~\ref{FNLduality}. Section~\ref{algebra} examines the fracton algebra, its extensions, and its connection to Carroll and Galilei algebras, offering both mathematical structure and physical interpretation. Section~\ref{Fracton Gravity and Non-Lorentzian Gravity} investigates the geometric realization of fracton algebra through gauging procedures and examines background geometries and curvature invariants arising from two distinct coupling schemes with fracton gauge fields, leading to the derivation of associated geometric constraint. Lastly, in Sectio~\ref{sec:COM}, we extend the Aristotelian algebra by introducing a center-of-mass operator and construct its geometric representation analogously to fractonic gauge field coupling. The paper concludes in Section \ref{conclusion} with a summary of key results and contributions and suggestions for future research directions.

\section{Fractons in gauge field} \label{Mathematical description of fractons}
In this section, we establish the canonical action for both mobile and immobile fractons, ensuring their invariance under fracton symmetries. This foundational work will be crucial for our subsequent discussions on Carrollian and Galilean particles and their duality with fractons. 
Fractons exhibit not only charge conservation, but also dipole moment conservation \cite{Pretko:2016kxt,Pretko:2016lgv,Pretko:2020cko}. From a mathematical perspective, fractons emerge as intriguing phenomena that can be described by tensor gauge theories. Applying the gauge principle ensures that this symmetry holds locally \cite{Pretko:2018jbi}. Specifically, fractons are characterized by a $U(1)$-valued symmetric tensor denoted as $A_{ij}$, with its conjugate variable $E_{ij}$ representing a generalized electric field. Remarkably, various rank-2 gauge transformations of $A_{ij}$ lead to analogues of Gauss’s law \cite{Pretko:2016kxt}:
\begin{align}
    \partial_i E^{ij}=\rho^j,\quad \partial_i \partial_j E^{ij}=\rho,\quad E^i_{i}=\rho_{tr}.
\end{align}
If we consider Gauss's law, $\partial_i \partial_j E^{ij}=\rho$, the electric charges in this theory (Scalar charge theory) are subject to two constraints \cite{Pretko:2016kxt,Pretko:2016lgv}:
\begin{align}
    \int \rho=\text{constant},\quad \int \vec x\rho= \text{constant},
\end{align}
corresponding to the conservation of charge and dipole moment. The dipolar bound states in this theory are fully mobile, exhibiting both longitudinal and transverse motion \cite{Pretko:2016kxt,Pretko:2016lgv}. By considering Gauss's law $\partial_i E^{ij}=\rho^j$ for a vector-valued charge $\rho^j$, the charges of this theory adhere to two constraints \cite{Pretko:2016kxt,Pretko:2016lgv}:
\begin{align}
    \int  \vec \rho=\text{constant},\quad \int \vec x \times \vec \rho= \text{constant},
\end{align}
demonstrate that both charge and angular momentum of vector charge remain conserved. To comply with these conservation laws, the vector charges are constrained to be one-dimensional particles. They can only move in the direction of their charge vector \cite{Pretko:2016lgv}.\\
By additionally imposing the tracelessness condition, $E^i_{i}=0$, the theory retains all conservation laws from the previous formulation and introduces two additional conservation laws associated with the tracelessness constraint \cite{Pretko:2016kxt,Pretko:2016lgv}:
\begin{align}
    \int  (\vec x.\vec \rho)=\text{constant},\quad \int [(\vec x . \vec \rho)\vec x- \frac{1}{2}x^2 \vec \rho]= \text{constant},
\end{align}
these conservation laws restrict the movement of fundamental charges, transforming them into fractons ($0$-dimensional particles). The only mobile particles in this theory are bound states \cite{Pretko:2016lgv}.

In general, the principle of charge conservation remains valid. Within the set of symmetries under consideration, the conserved quantity associated with the corresponding generator, denoted by $\hat{Q}$, characterizes the fracton monopole charge. Additionally, we introduce dipole symmetry, with its generator $\hat{Q}_i$ corresponding to the conserved dipole moment. The corresponding conservation laws for charge and dipole can be denoted as \cite{Pretko:2016kxt,Pretko:2016lgv,Grosvenor:2021hkn}
\begin{align} 
    & \frac{d\hat{Q}}{dt }=\frac{d}{dt}\int \extd ^dx \, \rho=0  , \label{two conservation laws 1} \\
   & \frac{d\hat{Q_i}}{dt }=\frac{d}{dt}\int \extd ^dx \; x_i\rho=0. \label{two conservation laws 2}
\end{align}
We infer from equations (\ref{two conservation laws 1}) and (\ref{two conservation laws 2}) that monopole charges, subject to the constraints of charge conservation and dipole symmetry, remain stationary \cite{Pretko:2016kxt,Pretko:2016lgv}. According to equation (\ref{two conservation laws 2}), we have $\dot{x}_i=0$. In addition to these two conservation laws, we assume that the system is translationally invariant, which implies the conservation of momentum
 \begin{align}
     \frac{d\hat{P_i}}{dt }=\frac{d{p_i}}{dt }=0.
 \end{align}
The generators $\hat{Q}$, $\hat{Q}_{i}$, $\hat{H}$, $\hat{J}_{ij}$ and $\hat{P}_{i}$ naturally generate the symmetry group referred to as the ``dipole'' or ``fracton'' symmetry, which has the following commutation relations \cite{Armas:2023ouk}:
\begin{equation} \label{dipole symmetry}
\begin{split}
    \{\hat{Q},\hat{Q}_{i}\} &= \{\hat{Q},\hat{P}_{i}\}=0,\quad \{\hat{P_i},\hat{Q}_{j}\}=\delta_{ij}\hat{Q} , \\
     \{\hat{P}_{i},\hat{J}_{jk}\} &=\delta_{ik}\hat{P}_{j}-\delta_{ij}\hat{P}_{k},\quad \{\hat{Q}_{i},\hat{J}_{jk}\}=\delta_{ik}\hat{Q}_{j}-\delta_{ij}\hat{Q}_{k}, \\
     \{\hat{J}_{ij},\hat{J}_{kl}\} &=\delta_{il}\hat{J}_{jk}-\delta_{ik}\hat{J}_{jl}+\delta_{jk}\hat{J}_{il}-\delta_{jl}\hat{J}_{ik},
     \end{split}
\end{equation}
where the Hamiltonian $\hat{H}$ is central and $\hat{J}_{ij}$ and $\hat{P_j}$ are rotation and momentum operators. 
According to equation (\ref{dipole symmetry}), all transformations preserve the electric charge of the system. However, based on this algebra, we identify the dipole symmetry as \cite{Grosvenor:2021rrt}
\begin{align} \label{dipole transformation 1}
\delta_{\beta} x_i=0,\quad \delta_{\beta} p_{i}=q\beta_i,\quad \delta_{\beta} E=0,\quad\delta_{\beta} t=0,
\end{align}
where $\beta_i$ is the parameter of dipole transformation and $E$ is the fracton energy.\\
The conservation of charge and dipole moments requires the existence of gauge fields $A_{\mu \nu}$ and a Maxwell-like theory for fractons, as demonstrated in \cite{Bertolini:2022ijb,Pretko:2016lgv}. In the following, we will formulate an action principle for fractons interacting with gauge fields. This action will be based on the Maxwell-like theories proposed for scalar and vector charge fractons. We will then derive the corresponding equations of motion and the Lorentz force.

It is worth highlighting the nontrivial interplay between translation symmetry ($ e^{i b_i \hat{P}_i}$ ) and dipole symmetry ( $e^{i \beta_i \hat{Q}_i}$ ), which typically emerge in fracton systems or in theories governed by higher-moment conservation laws. In infinite spaces with trivial topology, where both charge and dipole moment are unambiguously defined, these symmetries exhibit well-behaved structures. In such settings, dipole symmetry manifests as a continuous  \text{U}(1) internal symmetry.
However, in systems with nontrivial topology, such as those endowed with periodic boundary conditions:
i) A careful analysis is necessary to understand the structure and implementation of dipole symmetry.
ii) The symmetry algebra may be altered; in particular, translations and dipole transformations may fail to commute or retain their standard interpretation \cite{Burnell:2023fsr, Rudelius:2020kta}.
iii) This leads to a nontrivial deformation of the underlying symmetry algebra \cite{Shirley:2017suz, Luo:2022mrj}.
For instance, in the presence of periodic boundaries, dipole symmetry becomes discrete due to topological constraints. Accordingly, the dipole transformation parameter $\beta_i$ must be discretized, taking values as integer multiples of $2\pi / L$. This discretization ensures that the dipole symmetry remains consistent under translations by $L$, where $L$ denotes the size of the periodic domain.

\subsection{Scalar charge theory}\label{sec:scalartheory}
\paragraph{Isolated Fractons:} Isolated fractons are an intriguing subject in condensed matter physics due to their unique properties, including restricted mobility. As noted in \cite{Pretko:2016lgv}, fractons do not interact with gauge fields and are immobile. We will now formulate a description of these particles using the canonical action as follows:
\begin{align}\label{Isolated Fractons}
    S_{\text{Fracton}}=\int \extd \tau[-E\dot{t}+(\vec{p}-q\vec{\chi})\cdot \dot{\vec x}-\frac{e}{2}(E^2-\theta^2)],
\end{align}
which is invariant under dipole transformations \eqref{dipole transformation 1} and $\vec{\chi}$ is a Lagrange multiplayer   with transformation rule $\delta_{\beta}\chi_{i}=\beta_{i}$. The physical interpretation of \(\vec{\chi}\) establishes an additional constraint on isolated fractons, requiring that they remain immobile. In this canonical action, \( e \) serves as a Lagrange multiplier to ensure that the fracton energy \( E \) remains constant and equal to \( \theta \).
The equations of motion for this type of particles are
\begin{align}\label{E.O.Ms of Isolated Fractons}
    \dot{t}=-eE,\quad\dot{\vec x}=0,\quad \dot{E}=0,\quad\dot{\vec p}=q\dot{\vec \chi}.
\end{align}
Isolated fractons must be unaffected by any forces or particles; therefore, we have \(\dot{\vec{p}} = 0\), which implies that \(\dot{\vec{\chi}} = 0\).
These are isolated particles that are non-interacting with gauge fields and remain stationary. In the subsequence sections, we couple fracton dipoles or bound states of fractons to symmetric tensor gauge fields to study their dynamics in the presence of gauge fields.

\paragraph{Static Dipole:} In this part, we explore the bound states of the fractons. As suggested in \cite{Pretko:2016lgv}, when fractons interact and form dipoles, they may exhibit mobility. We will investigate these bound states under two scenarios: mobile dipoles and stationary dipoles. To describe stationary fracton dipoles, we employ the following action:
\begin{align}\label{Static Dipole}
 S_{\text{Dipole}}=\int \extd \tau [-E\dot{t}+\vec{p}\cdot \dot{\vec x}-\frac{e}{2}(E^2-\theta^2)+d^{i}\dot t\partial_{i}A_0(\vec x,t)-\dot{x}^{i}{d}^jA_{ij}(\vec x,t)] .
\end{align}
This action is invariant under dipole transformations, where $d^{i}$ represents the dipole vector. Just as ordinary charged particles generate and interact with gauge fields, we extend this concept to fracton dipoles. We will explore their motion under the influence of the gauge fields $A_0$ and $A_{ij}$, with the following transformation rules \footnote{More details for the Maxwell-like theories of these gauge fields can be find in \cite{Bertolini:2022ijb,Pretko:2016lgv}},
\begin{align} \label{action:DS}
    \delta_{\beta}p_{i}=d^{j}\beta_{ij},\quad \delta_{\beta}A_{ij}=\beta_{ij},\quad\delta_{\beta}E=\delta_{\beta}\vec{x}=\delta_{\beta}t=\delta_{\beta}A_{0}=0.
\end{align}
Consequently, we can express the equations of motion as follows:
\begin{equation}\label{fracton e.o.m 1}
\begin{split}
    \dot{t}&=-eE,\quad\dot{\vec x}=0,\\
    \dot{E}&=-d^{i}\partial_{i}\dot{A}_{0}+d^{i}\dot{t}\partial_{t}\partial_{i}A_{0}- \dot{x}^{j}d^{i}\partial_{t}A_{ij} = 0, \\
    \dot{p}_{m}&=d^{j}\dot{A}_{mj}+d^{i}\dot{t}\partial_{i}\partial_{m}A_{0}(\vec x,t)-\dot{ x}^{i}d^{j}\partial_{m}A_{ij}
    \end{split}
\end{equation}
where in the second line we expanded the total time derivative  $\dot{A}_{0}$. The equations of motions (E.O.Ms) derived from the action \eqref{Static Dipole} demonstrate that the dipole $d^i$ is stationary while experiencing a force from the fractonic electromagnetic fields. Using the E.O.Ms, through gauge fixing we can make $E=cte$.
The equation for $\dot{p}_{i}$ reveals that the Lorentz force acting on a fracton dipole can be expressed as follows:
\begin{align}
    \dot{p}_{m}=d^{j}(\dot{x}^{i}\partial_{i}A_{mj}+\dot{t}\partial_{t}A_{mj})+\dot{t}d^{i}\partial_{i}\partial_{m}A_{0}(\vec x,t)-\dot{ x}^{i}d^{j}\partial_{m}A_{ij},
\end{align}
according to the gauge theory of fractons \cite{Pretko:2016lgv}, the fracton electric field can be defined as
\begin{align*}
    E_{jm}=\partial_{t}A_{jm}+\partial_{j}\partial_{m}A_{0},
\end{align*}
in which $E_{jm}$ represents a generalized electric field of fractons.
Since fractons are immobile, i.e., $\dot{\vec x}=0$, by fixing the gauge $t= \tau$, the only force contributing to the Lorentz force is the electrostatic force, which we can express as follows:
\begin{align}
    \dot{p}_{m}=d^{j}\partial_{t}A_{mj}+d^{j}\partial_{j}\partial_{m}A_{0}=d^{j}E_{mj}.
\end{align}
\paragraph{Mobile Dipole:} In the scalar charge theory, we discovered that bound states of charges can form dipoles. We can identify two distinct types of dipoles within this theory. In Section 6, we will directly relate these two models of fracton dipoles to other non-Lorentzian theoretical particles that exist within different symmetries.

The first mobile dipole can be constructed in this theory, has the following the canonical action,
\begin{align}\label{Mobile Dipole 1}
    S_{\text{Dipole}}=\int \extd \tau [-E\dot{t}+\vec{p}\cdot \dot{\vec x}-\frac{e}{2}(p^2-\epsilon^2)+\vec{\chi}\cdot(\vec{d}-\vec{\psi})+d^{i}\dot t\partial_{i}A_0(\vec x,t)-\dot{x}^{i}{d}^jA_{ij}(\vec x,t)].
\end{align}
To ensure the action's invariance under dipole transformations, we introduced an additional term with a Lagrange multiplier, $\vec{\chi}$. In the above action, the Lagrange multiplier $\vec{\chi}$ should transform under dipole transformations as $ep^{j}\beta_{ij}$. As a consequence of this additional term, the dipole moment $\vec{d}$ becomes a constant, which we have fixed to $\vec{\psi}$. The resulting equations of motion are:
\begin{equation}\label{E.O.Ms Mobile Dipole 1}
\begin{split}
    \dot{t}&=0,\quad\dot{\vec x}=e\vec p,\\
    \dot{E}&=-d^{i}\partial_{i}\dot{A}_{0}+d^{i}\dot{t}\partial_{t}\partial_{i}A_{0}- \dot{x}^{j}d^{i}\partial_{t}A_{ij}=-2d^{i}\dot{x}^{j}E_{ij},\\
    \dot{p}_{m}&=d^{j}\dot{A}_{mj}+d^{i}\dot{t}\partial_{i}\partial_{m}A_{0}(\vec x,t)-\dot{x}^{i}d^{j}\partial_{m}A_{ij}=d^{j}\dot{x}^{i}(\partial_{i}A_{mj}-\partial_{m}A_{ij}).
    \end{split}
\end{equation}
These equations of motion demonstrate that the dipole fracton bound states can move under an external fractonic electromagnetic force. However, in this model of mobile dipoles, the Lorentz force includes only the magnetic component because, in this scenario, we have $\dot{t} = 0$. Therefore, the Lorentz force exerted on the fracton dipole will be
\begin{align}\label{diple:lorentz}
    \dot{p}_{m}=d^{j}(\epsilon_{ijl}u^{i}B_{m}^{l}),
\end{align}
in which the fractonic magnetic field is defined as $ B_{m}^{l}=\epsilon^{ijl}\partial_{i}A_{mj}$ \cite{Pretko:2016lgv}. 
Eq.~\eqref{diple:lorentz} first proposed in \cite{Pretko:2016lgv} and then derived from completely different approach in \cite{Bertolini:2022ijb}, here we give an action principle description for this theory, completing the picture for the fracton dipoles and their motion.

A second model for mobile dipoles can be expressed in the canonical form, as shown below:
\begin{align}\label{Mobile Dipole 2}
    S_{\text{Dipole}}=\int d \tau [-E\dot{t}+\vec{p}\cdot \dot{\vec x}-\frac{e}{2}(p^2-\epsilon^2)-\frac{\widetilde e}{2}(E^2-\theta^2)+\vec{\chi}\cdot(\vec{d}-\vec{\psi})+d^{i}\dot t\partial_{i}A_0(\vec x,t)-\dot{x}^{i}{d}^jA_{ij}(\vec x,t)].
\end{align}
To ensure invariance under dipole transformations, we introduced an additional term with a Lagrange multiplier $\widetilde e$, which ensures the energy of the particles is constant. The resulting equations of motion are:
\begin{equation}\label{E.O.Ms Mobile Dipole 2}
\begin{split}
    \dot{t}&=-eE,\quad\dot{\vec x}=e\vec p, \\
    \dot{E}&=-d^{i}\partial_{i}\dot{A}_{0}+d^{i}\dot{t}\partial_{t}\partial_{i}A_{0}- \dot{x}^{j}d^{i}\partial_{t}A_{ij} =-2d^{i}\dot{x}^{j}E_{ij},\\
    \dot{p}_{m}&=d^{j}\dot{A}_{mj}-d^{i}\dot{t}\partial_{i}\partial_{m}A_{0}(\vec x,t)-\dot{x}^{i}d^{j}\partial_{m}A_{ij}=d^{j}(E_{mj}+\epsilon_{ijl}\dot{x}^{i}B_{m}^{l}).
    \end{split}
\end{equation}
These equations of motion demonstrate that the fracton dipole bound states can move under an external fractonic electromagnetic force. However, in this model of mobile dipoles, the Lorentz force includes both electric and magnetic components, which is
\begin{align}
    \dot{p}_{m}=d^{j}(E_{mj}+\epsilon_{ijl}\dot{x}^{i}B_{m}^{l}).
\end{align}

\subsection{Vector charge theory}\label{section:vector-charge}
The study of vector charges in fracton theories is motivated by the desire to explore a wider range of mobility restrictions and topological phases. Vector charges introduce additional degrees of freedom, allowing for more complex patterns of restricted motion. This provides researchers with a valuable tool for understanding and manipulating these exotic excitations, connecting fracton theories to other fields of physics, and complementing the study of scalar charges. By considering vector charges, we can uncover new and exciting phenomena in condensed matter physics and particle physics at extreme limits.\\
In the previous subsection, we established that isolated fractons are immobile particles; however, they can form dipoles that can move in any direction. 
Vector charge theory allows us to describe a new type of particle known as lineons, which exhibit restricted mobility along their own vector charge axis and are invariant under dipole transformations; their gauge theory is developed in \cite{Pretko:2016lgv}.
The canonical action describing fractonic vector charges with restricted mobility is as follows:
\begin{align}\label{action:lineon}
    S_{\text{Lineons}}=\int \extd \tau [-E\dot{t}+\vec{p}\cdot \dot{\vec x}-\frac{\widetilde e}{2}(E^2-\theta^2)-\frac{e}{2}({p}^2-{\epsilon}^2)+\vec{\chi}\cdot(\vec{q}-\vec{\psi})+q^{i}\dot t\phi_{i}(\vec x,t)-\dot{x}^{i}{q}^jA_{ij}(\vec x,t)].
\end{align}
This action is also invariant under the vector charge transformations. The vector charge transformations for fractonic vector charge theory are:
\begin{align}
    \delta_{\beta}p_{i}=q^{j}\beta_{ij},\quad \delta_{\beta}A_{ij}=\beta_{ij},\quad\delta_{\beta}E=\delta_{\beta}\vec{x}=\delta_{\beta}t=\delta_{\beta}A_{0}=0.
\end{align}
These symmetrical transformations can be expressed using vector charge algebra, as discussed in section (\ref{algebra}), particularly in (\ref{vector charge algebra}). 

Similarly to the mobile dipole in scalar charge theory, to ensure the action's invariance, we must introduce a Lagrange multiplier. In action \eqref{action:lineon}, the Lagrange multiplier $\vec{\chi}$ should transform under dipole transformations as $ep^{j}\beta_{ij}$. This transformation guarantees the action's invariance under dipole transformations. Consequently, the equations of motion derived from this action describe the dynamics of a system with a fractonic vector charge, given by:
\begin{equation}\label{e.o.ms of lineons}
\begin{split}
    \dot{t}&=-eE,\quad\dot{\vec x}=e\vec p ,\quad
    \vec q=\vec \psi,\\
    \dot{E}&=-q^{i}\dot{\phi}_{i} +q^{i}\dot{t}\partial_{t}\phi_{i}-q^{i}\dot{x}^{j}\partial_{t}A_{ij}=-q^{i}\dot{x}^{j}(\partial_{j}\phi_{i}+\partial_{t}A_{ij})\ =q^{i}\dot{x}^{j}E_{ij}\\
    \dot{p}_{m}&=q^{j}\dot{A}_{mj}-q^{i}\dot{t}\partial_{m}\phi_{i}(\vec x,t)-\dot{\vec x}^{i}q^{j}\partial_{m}A_{ij}=q^{j}[\dot{x}^{i}(\epsilon^{ijl}\partial_{i}A_{mj})+E_{jm}]\\&=q^{j}(E_{jm}+\epsilon_{ijl}\dot{ x}^{i}B_{m}^{l}).
    \end{split}
\end{equation}
where the fractonic electric field is defined as \cite{Manoj:2020bcz}
    \begin{align}
        E_{jm}=-\partial_{m}\phi_{j}-\partial_{t}A_{jm}=-\frac{1}{2}(\partial_{m}\phi_{j}+\partial_{j}\phi_{m})-\partial_{t}A_{jm}\, .
    \end{align}
As previously mentioned, this theory includes additional conservation laws. One such law governs the moment of the vector charge, expressed as follows:
\begin{align}
\int  \vec x \times \vec \rho=\text{constant}.
\end{align}
A consequence of this conservation law is that an isolated charge can only move along its own vector charge $\vec{q}$. Therefore, these one-dimensional particles, called lineons, experience an effective electric field but are unaffected by magnetic fields, as their one-dimensional trajectories cannot enclose any magnetic flux. This leads to a simplified Lorentz force law:
\begin{align}\label{force-lineon2}
   |\textbf{F}|:= F_{j}q^j =E_{ik} q^{j}q^{k}.
\end{align}
It is important to note that force \eqref{force-lineon2} aligns with expectations made in \cite{Pretko:2016lgv} based on physical assumptions. In this work, we have derived this result from an action principle for lineons.

\section{Non-Lorentzian  point particle} \label{Non-Relativistic point particle}
In physics, the dynamics of a system can be described in two fundamentally equivalent ways: through equations of motion or through an action principle. Action is a scalar quantity that characterizes how a physical system changes dynamically. The equations of motion for the system can be derived by varying the action with respect to the dynamical variables. In this section, we construct a framework for non-relativistic particle theories by systematically applying limiting procedures to a relativistic particle model in two distinct extreme regimes. Through this approach—filling a gap in the existing literature—we derive the corresponding equations of motion and identify Time-like Galilean and Space-like Carroll particles as limiting cases of a relativistic theory. This method stands in contrast to previous works, which employed seed Lagrangians to construct the associated canonical actions for these particles \cite{Bergshoeff:2022qkx}.
\subsection{Relativistic point particle} \label{Relativistic point particle  and Poincare algebra}
In this subsection, we will briefly review the dynamics of relativistic point particles, which will serve as a foundation for studying this theory at its limits in the following sections. Throughout our analysis, we utilize the canonical action to derive symmetries and equations of motion for particles.\\
We work in flat $D$-dimentional space-time with the signature (+ - - -), where we have $D=1+d$ in which d is spatial dimension.
The canonical form of the action of a massive relativistic point particle is given by
\begin{align} \label{RPPA}
     S=\int\extd\tau L(x^\mu,p^\mu,e;\dot x^\mu,\dot p^\mu)=\int\extd\tau[-p_\mu\dot x^\mu- \frac{e}{2}(p^2-m^2)],
\end{align}
where $x^\mu$, $p^\mu$ and $e$  are the space-time coordinates,  their momentum and the einbein variables, respectively.  
The dot denotes the differentiation with respect to the parameter $\tau$. This action is invariant under the Poincaré transformation, as follows:
\begin{align} \label{ Poincare trans}
    \delta_Px^\mu=\Lambda_\nu^\mu x^\nu+a^\mu,\quad \delta_Pp^\mu=\Lambda_\nu^\mu p^\nu,\quad \delta_Pe=0.
\end{align}
To derive the equation of motions, we vary the action  (\ref{RPPA}) in terms of dynamical variables, $\delta S= \delta \int\extd\tau L(x^\mu,p^\mu,e;\dot x^\mu,\dot p^\mu)$, which results in the following Euler-Lagrange equations,
\begin{equation}\label{e.o.m 1}
\begin{split}
    & \int\extd\tau (\frac{\delta L}{\delta x^\mu}-\partial_\tau \frac{\delta L}{\delta \dot x^\mu})\delta x^\mu+\int d\tau \partial_\tau (\frac{\delta L}{\delta \dot x^\mu}\delta x^\mu)=0 ,\\
         & \int d\tau (\frac{\delta L}{\delta p^\mu}-\partial_\tau \frac{\delta L}{\delta \dot p^\mu})\delta p^\mu+\int d\tau \partial_\tau (\frac{\delta L}{\delta \dot p^\mu}\delta p^\mu)=0 ,
    \end{split}
    \end{equation}
then the equation of motions of the canonical action (\ref{RPPA}) are:
\begin{align} \label{e.o.m 2}
    \dot x^\mu=-ep^\mu,\quad \dot p^\mu=0 \, ,
\end{align}
with the following primary constraints,
\begin{align} \label{primary 1}
    \pi_{x^\mu}=-p^\mu,\quad \pi_{p^\mu}=0,\quad \pi _e=0.
\end{align}
Since the canonical Lagrangian in (\ref{RPPA}) is invariant under the Poincare transformation (\ref{ Poincare trans}), according to the Noether theorem, we can define the following Noether current or the conserved current associated with the Poincare transformation (\ref{ Poincare trans}),
\begin{align} \label{Noether 1}
    J_p^\tau=\frac{\delta L}{\delta \dot x^\mu}\delta_Px^\mu+\frac{\delta L}{\delta \dot p^\mu}\delta_Pp^\mu.
\end{align}
Every symmetry transformation has symmetry generators associated with the Noether current. 
In this case, the symmetry generators associated with the Noether current (\ref{Noether 1}) are called Poincare generators,
\begin{align}
     \hat{P_\mu}= p_\mu=\partial_\mu,\quad \hat{J}_{\mu\nu}=(x_\mu p_\nu-x_\nu p_\mu),
\end{align}
and the corresponding algebra, which is related to these operators, is Poincare algebra, as follows:
\begin{equation} \label{Poincare algebra}
\begin{split}
    \{\hat{P_\mu},\hat{P_\nu}  \} &=0,\quad \{\hat{J}_{\mu \nu},\hat{P_\rho} \}=-(\eta_{\mu \rho} \hat{P_\nu}-\eta_{\nu \rho} \hat{P_\mu}), \\
    \{\hat{J}_{\mu \nu},\hat{J}_{\rho \sigma}\} &=\eta_{\nu \rho}\hat{J}_{\mu \sigma}+\eta_{\mu \sigma}\hat{J}_{\nu \rho}-\eta_{\mu \rho}\hat{J}_{\nu \sigma}-\eta_{\nu \sigma}\hat{J}_{\mu \rho}.
    \end{split}
\end{equation}
\subsection{Galilean theory}\label{Galileantheory}
In daily life, we often observe phenomena that are not governed by Lorentzian symmetry. Instead, these phenomena conform to Galilean symmetry, a non-relativistic framework relevant for classical mechanics. This symmetry applies to objects whose motion is significantly slower than light's speed. Galilean symmetry is especially crucial in condensed matter physics, where quantum effects are prominent, but the velocities are non-relativistic. The Galilean symmetries can in fact be seen as the limit of Lorentz or Poincaré symmetries when the speed of light is considered to be infinitely large ($c\rightarrow \infty$). In this limit, time is absolute in the Galilean framework and is the same for all observers, regardless of their relative motion. 
\subsubsection{Galilean transformation}
Galilean transformations belong to the Galilean group, encompassing spatial rotations, spatial and temporal translations, and Galilean boosts. These transformations can be derived from Poincaré transformations by applying a specific limit. The Poincaré transformations dictate the space-time coordinate changes as follows \cite{levy1965nouvelle}:
\begin{align} \label{poincare trans2 }
\begin{split}
        x^{\prime \mu}=\Lambda^\mu_\nu x^\nu-a^\mu\quad\quad\quad\quad\\ x^{\prime 0}=ct^{\prime}=\Lambda^0_\nu x^\nu -a^0 \quad\quad\\ \vec x^{\prime}=\Lambda^i_\nu x^\nu+\vec a, (i=1,2,3),
\end{split}
\end{align}
from which, the Lorentz transformations are:
\begin{align}\label{poincare trans3}
ct^{\prime}=\gamma_{\zeta}(ct-\vec \zeta.\vec x),\quad \vec x^{\prime}=\gamma_{\zeta}(\vec x-\vec \zeta ct),\quad \gamma_{\zeta}=\frac{1}{\sqrt{1-\vec \zeta^{\,2}}},
\end{align}
where $\vec \zeta$ is the Lorentz boost transformation parameter. To obtain the Galilean transformations, we apply the scaling limit $\vec \zeta=\dfrac{-\vec \beta_{G}}{c}$, in which the parameter $\vec \beta_{G}$ represents the Galilean boost parameter, then we get:
\begin{align}
    t^{\prime}=t+a_0,\quad \vec x^{\prime}= R\vec x-\vec \beta_{G} t+ \vec a.
\end{align}
The key difference between the Galilean transformations and the Poincaré transformations is that the Galilean transformations assume an absolute time that is the same for all observers, regardless of their motion, and a space where distances are absolute. Now we can derived the Galilean transformation for energy and momentum by taking a limit on the Lorentz transformation on energy and momentum. In relativistic regime we have:
\begin{align}\label{poincare momen}
    {\vec p}^{\, \prime}=\gamma_{\zeta}(\vec p-\vec \zeta \frac{E}{c}),\quad \frac{E^{\prime}}{c}=\gamma_\zeta(\frac{E}{c}-\vec \zeta.\vec p),
\end{align}    
using the limit $c\rightarrow \infty$ we get the following relations:
\begin{align}
    {\vec p}^{\, \prime}=\vec p,\quad E^{\prime}=-\vec \beta_{G}.\vec p+E .
\end{align}
In general, the Galilean transformations can be summarized as follows \cite{levy1971galilei,castellani1998galilean}:
\begin{align} \label{Galilean trans}
    \delta_Gt=a_t,\quad \delta_G \vec x=R\vec x-\vec \beta_{G} t+\vec a,\quad \delta_G\vec p=R\vec p,\quad \delta_G E=-\vec \beta_{G}.\vec p.
\end{align}
$ \delta_G$ stands for Galilean variations. We see that there are no boost transformations in time, which indicates that we have absolute time in Galilean symmetry. Additionally, under a Galilean boost, momentum remains unchanged.
\subsubsection{Massless Galilean particles}\label{Massless Galilean particles}
Particles exhibiting Galilean symmetry are termed Galilean particles. This section explores their dynamics, noting that while the properties of massive Galilean particles are well documented in the classical and quantum realms, the concept of massless Galilean particles remains largely unexplored, yet their existence is plausible. The action of a free non-relativistic particle is invariant under the Galilean transformation, with the particle's mass emerging as a central charge in the algebra of Noether charges, known as the Bargmann algebra \cite{Figueroa-OFarrill:2024ocf,Batlle:2017cfa}.\\
A Galilean dynamical system that embodies the Galilei algebra without the central charge is known as a ``massless Galilean'' system; for further details please see \cite{souriau1970structure}. The massless Galilean particle may be considered as a non-relativistic counterpart to the massless relativistic particle, which can be constructed and obtained by limiting process when $c \rightarrow \infty$. Authors of \cite{souriau1970structure} suggest distinct non-relativistic particles, each labeled by a parameter termed `color' ($k$ ), which uniquely identifies each theoretical particle type\footnote{
Thus, without the need to consider relativistic mass-energy equivalence, color $k$ becomes a useful tool for categorizing these particles and exploring their properties and interactions \cite{souriau1970structure}}. Gomis et al. have established that a massless Galilean particle characterized by the color parameter ($k$) represents a non-relativistic limit case of a relativistic tachyon \cite{Batlle:2017cfa}. In this study, we restrict our scope to cases where ($k=0$), thereby focusing exclusively on massless relativistic particles that are not tachyons \footnote{A discussion on $k=0$ and massless Galilean particles}. \\ %
Under Galilean symmetry, we distinguish between space-like and time-like particles based on their behavior during Galilean transformations. The distinction between these two categories may be attributed to their respective dynamics when the spatial variations of the field are significantly different from its temporal changes or vice versa.\\
We know that the Galilean symmetry is the result of the limit where $c$ approaches infinity. Rather than utilizing a dimensionfull parameter such as the speed of light to approach infinity, it is preferable to use a dimensionless parameter for limiting purposes. There is a possible way to achieve this by rescaling these parameters using dimensionless parameters and considering limiting dimensionless parameters approaching zero or infinity.
Now we introduce the following scaling,
\begin{equation*} 
     c\rightarrow \omega c, \quad \quad \omega\rightarrow \infty . 
\end{equation*}
To attain the two Galilean limits, we can adjust the scale of the canonical variables.
The space-like limit, which is the first type of Galilean limit, is reached by rescaling only the spatial components of spatial coordinates, momentum and electric charge of particles as follows
\begin{align} \label{Galelian first type lim}
    \vec x\rightarrow \frac{ \vec x}{\omega},\quad \vec p \rightarrow {\omega}{\vec p},\quad e\rightarrow -\frac{ e}{\omega^2} \, .
\end{align}
Conversely, the time-like limit, which is the second type of Galilean limit, is achieved through rescaling of the temporal components of time coordinate, energy and electric charge of particles
\begin{align} \label{Galelian second type lim}
    x^0\rightarrow \frac{ t}{\omega},\quad p^0 \rightarrow {\omega}{E},\quad e\rightarrow+\frac{ e}{\omega^2} \, .
\end{align}
By applying the two limits, denoted as (\ref{Galelian first type lim}) and (\ref{Galelian second type lim}), to the canonical action (\ref{RPPA}), we derive two distinct actions that describe Galilean particles. These actions correspond to two separate classes of Galilean particles: time-like and space-like. In the context of electromagnetism, particularly regarding the electric charge, these are often referred to as the electric and magnetic limits of Galilean electromagnetism, as emphasized by Le Bellac and Lévy-Leblond \cite{LeBellac:1973unm,rousseaux2013forty}. However, to avoid potential confusion when considering charged particles in electromagnetic fields, we will use a different terminology in this work. The derivation of these two unique canonical actions for Galilean particles will be elaborated in the following subsections. 
\subsubsection{Space-like Galilean Particles}
To enhance clarity and ensure self-containment, we will rederive the theory for space-like Galilean particles. We begin with the canonical action of relativistic particles, Eq.~\eqref{RPPA}, considering two possible routes for taking the Galilean limit: Eq.~\eqref{Galelian first type lim} and Eq.~\eqref{Galelian second type lim}. We first adopt the limit specified in Eq.~\eqref{Galelian first type lim}, accompanied by the rescaling $m \rightarrow \frac{\kappa}{\omega}$ and the limit $\omega \rightarrow \infty$. Under these transformations, the canonical action yields \footnote{In our approach, we derived this action without color $k$. One of the key points of our work is deriving this action directly from its relativistic counterparts with their details.}:
\begin{align} \label{magnetic action}
S=\int d\tau L_{mG}(t,E,\vec x,\vec p;\dot t,\dot E,\dot {\vec x},\dot {\vec p})=\int d\tau[-E\dot t+\vec {p} .\dot{\vec{x}}-\frac{e}{2}(\vec p^{\ 2} + {\kappa}^{2})].
\end{align}
where ${\kappa}$ is a constant that fixes the momentum, and can be related to the so-called 'colour' parameter, since in Galilean dynamics mass parameter can not be defined, as discussed in \cite{souriau1997structure}. This action is directly derived from their massless relativistic counterparts using the limit method. In \cite{Batlle:2017cfa} and \cite{Bergshoeff:2022qkx}, the massless Galilean action with color $k$ is derived based on Souriau's method. In this paper, we derive this action directly from relativistic massless theory, bypassing Souriau's method and instead utilizing a limiting procedure.
The canonical action (\ref{magnetic action}) is invariant under the Galilean transformation (\ref{Galilean trans}) as follows:
\begin{align*} 
    \delta_Gt=a_t,\quad \delta_G \vec x=R\vec x-\vec \beta t+\vec a,\quad \delta_G\vec p=R\vec p,\quad \delta_G E=-\vec \beta.\vec p \, ,
\end{align*}
if the Galilean transformation of the canonical variable $e$ is given by:
\begin{align} \label{Galilean trans 2}
    \delta_G e=0.
\end{align}
The dynamics of a system can be described by equations of motion derived from the Lagrangian (\ref{magnetic action}) by varying its canonical variables. Then, the equations of motion from the canonical action (\ref{magnetic action}) are:
\begin{align} \label{space-like Gal e.o.m}
    \dot t=0, \quad \dot E=0,\quad \dot {\vec x}=e\vec p,\quad \dot{\vec p}=0.
\end{align}
The canonical Lagrangian in (\ref{magnetic action}) is invariant under the Galilean transformations (\ref{Galilean trans}) and (\ref{Galilean trans 2}), as we noted earlier, the existence of symmetry transformations imply the existence of a conservation law. According to the Noether theorem, for every continuous symmetry, there is a corresponding conserved current.
Then the Noether current associated with the Galilean transformations (\ref{Galilean trans}) and (\ref{Galilean trans 2}), is as follows:
\begin{align}\label{Gal Noether 2}
    J_G^\tau=\frac{\delta L_{mG}}{\delta \dot t} \delta_Gt+ \frac{\delta L_{mG}}{\delta \dot E} \delta_GE+\frac{\delta L_{mG}}{\delta \dot {\vec x}} \delta_G \vec x+\frac{\delta L_{mG}}{\delta \dot {\vec p}} \delta_G \vec p.
\end{align}
The symmetry generators associated with the Noether current (\ref{Gal Noether 2}) are called Galilean generators, which can be listed as:
\begin{align} \label{Galilean operators}
    \hat{H}=E,\quad \hat{P}=\vec p ,\quad \hat{B}=\vec p t , \quad \hat{J}=\vec x  \times \vec p .
\end{align}
This theory has the following primary constraints,
\begin{align} \label{Gal primery 2}
    \pi_E=0,\quad \pi_t=-E,\quad \pi_{\vec{x}}=\vec p,\quad \pi_{\vec{p}}=0,\quad \pi_e=0,
\end{align}
where we denoted the canonical momenta of the variables by $\pi$. The following Poisson brackets can be derived from  the canonical action (\ref{magnetic action}):
\begin{align}
    \{\ E,\ t\}=1, \quad \{\ e,\pi_e\}=1,\quad \{\ x^i, \ p^j\}=\delta^{ij} \quad i,j=1,2,3 \, .
\end{align}
Then the corresponding algebra of generators (\ref{Galilean operators}), is the Galilean algebra, 
\begin{equation}
\begin{split}
    \{\ E,\hat{P}_{i}\} &=\{\ E,\hat{J}_{ij}\}= \{\hat{P}_{i}.\hat{B}_{j}\}=0 , \\
      \ \{\ E,\hat{B}_{i}\} &= \hat{P}_{i},\quad
    \{\hat{P}_{i},\hat{J}_{jk}\}=-\hat{P_k} \delta_{ij}+\hat{P}_{j}\delta_{ik},\quad \{\hat{B}_{i},\hat{J}_{jk}\}=-\hat{B}_{k}\delta_{ij}+\hat{B}_{j}\delta_{ik}, \\
    \{\hat{J}_{ij},\hat{J}_{kl}\} &=\delta_{ik}\hat{J_{jl}}-\delta_{il}\hat{J_{jk}}-\delta_{jk}\hat{J}_{il}+\delta_{jl}\hat{J}_{ik}.
        \end{split}
\end{equation}
\subsubsection{Time-like Galilean Particles}
In this subsection, we derive the theory of time-like massless Galilean particles through a limiting procedure, departing from previous studies that relied on a seed Lagrangian approach \cite{Bergshoeff:2022qkx}. In the preceding subsection, we obtained space-like Galilean particles by applying the first Galilean limit \eqref{Galelian first type lim} to the canonical action (\ref{RPPA}). Now, by employing the second Galilean limit \eqref{Galelian second type lim}, together with the rescaling $m \rightarrow \frac{\epsilon}{\omega}$, we obtain the corresponding action for Time-like Galilean particles, given by:
\begin{align} \label{first elec Galilean action}
    S=\int d\tau L=\int d\tau[-E\dot t+\vec p.\dot{\vec x}-\frac{e}{2}(E^2 - \epsilon^2)].
\end{align}
However, this action is not invariant under the Galilean transformations (\ref{Galilean trans}); because of the existence of last term $\dfrac{e}{2}(E^2 - \epsilon^2)$, which transforms as:
\begin{align}
\delta_G(E^2 - \epsilon^2)= -2{\Vec{p}}.\vec \beta E.
\end{align}
A similar relation can also be found in \cite{Bergshoeff:2022qkx}.
We can include a Lagrange multiplier term in order to make the canonical action (\ref{first elec Galilean action}) invariant with respect to the Galilean transformations (\ref{Galilean trans}) and (\ref{Galilean trans 2}). Then we have the following action:
\begin{align} \label{second elec Galilean action}
    S=\int d\tau L_{eG}(t,E,\vec x,\vec p;\dot t,\dot E,\dot {\vec x},\dot {\vec p}) =\int d\tau[-E\dot t+\vec{p}.\dot{\vec{x}}-\frac{e}{2}(E^2 - \epsilon^2)+\vec \chi.\vec p] , 
\end{align}
which is in agreement with the action derived from a different approach and reported in \cite{Bergshoeff:2022qkx}.
In the time-like limit for Galilean massless particles, the focus is on the temporal evolution of the particles, which is consistent across different inertial frames since time is absolute in Galilean relativity.
If we apply the Galilean transformations on this new variable, $\vec \chi$, we find:
\begin{align} \label{Galilean trans 3}
    \delta_G \vec \chi =-eE\vec{\beta}.
\end{align}
To derive the equations of motion, we vary the action (\ref{second elec Galilean action}) w.r.t the dynamical variables, from which we obtain:
\begin{align} \label{e.o.m 4}
    \dot E=0,\quad \dot{\vec x}=-\vec \chi,\quad \dot t=-eE,\quad \dot{\vec p}=0,\quad \vec p=0.
\end{align}
To understand this equation of motion, it can be inferred that Galilean time-like particles exhibit quasi-static behavior, since setting $\vec{\chi}=0$ is not feasible, since its absence breaks the Galilean symmetry.\\
Now the canonical Lagrangian (\ref{second elec Galilean action}) is invariant under the Galilean transformations (\ref{Galilean trans}), (\ref{Galilean trans 2}) and (\ref{Galilean trans 3}) . Therefore, the Galilean transformations are symmetry transformations of the canonical Lagrangian (\ref{second elec Galilean action}). Then the Noether current or the conserved current associated with the Galilean transformations (\ref{Galilean trans}), (\ref{Galilean trans 2}) and (\ref{Galilean trans 3}) is as follows:
\begin{align}
    J_G^\tau=\frac{\delta L_{eG}}{\delta \dot t} \delta_Gt+ \frac{\delta L_{eG}}{\delta \dot E} \delta_GE+\frac{\delta L_{eG}}{\delta \dot {\vec x}} \delta_G \vec x+\frac{\delta L_{eG}}{\delta \dot {\vec p}} \delta_G \vec p+\frac{\delta L_{eG}}{\delta \dot {\chi}} \delta_G\chi.
\end{align}
The symmetry generators are the same as the space-like limit symmetry generators in (\ref{Galilean operators}). Therefore, we have:
\begin{align} 
    \hat{H}=E,\quad \hat{P}=\vec p ,\quad \hat{B}=\vec p t , \quad \hat{J}=\vec x  \times \vec p.
\end{align}
By denoting the canonical momenta of the variables as $\pi$, we have the following primary constraints:
\begin{align}\label{Gal primary 3}
    \pi_E=0,\quad \pi_t=-E,\quad \pi_{\vec{x}}=\vec p,\quad \pi_{\vec{p}}=0,\quad \pi_e=0,\quad \pi_\chi=0.
\end{align}
Finally we will again obtain the Galilean algebra in the  time-like limit:
\begin{equation}
\begin{split}
    \{\ E,\hat{P}_{i}\} &=\{\ E,\hat{J}_{ij}\}= \{\hat{P}_{i}.\hat{B}_{j}\}=0 , \\
      \ \{\ E,\hat{B}_{i}\} &= \hat{P}_{i},\quad
    \{\hat{P}_{i},\hat{J}_{jk}\}=-\hat{P_k} \delta_{ij}+\hat{P}_{j}\delta_{ik},\quad \{\hat{B}_{i},\hat{J}_{jk}\}=-\hat{B}_{k}\delta_{ij}+\hat{B}_{j}\delta_{ik}, \\
    \{\hat{J}_{ij},\hat{J}_{kl}\} &=\delta_{ik}\hat{J_{jl}}-\delta_{il}\hat{J_{jk}}-\delta_{jk}\hat{J}_{il}+\delta_{jl}\hat{J}_{ik}.
        \end{split}
\end{equation}

\subsection{Carrollian Theory} \label{Carroll sym}
The Galilean symmetries emerge from the non-relativistic limit of Lorentz symmetries, where the speed of light $c$ tends towards infinity. In this study, we also explore the scenario in which the speed of light approaches zero, giving rise to Carroll symmetries, as detailed in \cite{Baiguera:2022lsw}. Contrary to non-relativistic symmetry, Carroll symmetry represents the ultra-relativistic limit of relativistic theories, as discussed in \cite{Bergshoeff:2014jla}
. The Carroll symmetry was originally understood by Lévy-Leblond in 1965 \cite{hansen2021beyond}, in Lévy-Leblond's own words, ``this symmetry is the degenerate cousin of Poincaré symmetry'' \cite{levy1965nouvelle,Duval:2014uoa}. We can say that the Carroll symmetry follows from the Poincare symmetry at the vanishing speed limit \cite{de2022carroll}. 
Carroll particles are theoretical entities that arise in the context of Carroll symmetry. Understanding Carroll particles involves examining their dynamics and how they relate to their relativistic counterparts. In a universe governed by Carroll symmetries, a single free Carroll particle exhibits no non-trivial dynamics. To gain a deeper understanding of Carroll particles, we will explore them in more detail in the following sections. We will begin by studying a free relativistic particle and then taking the limit
\begin{align} \label{vanishing c}
     c\rightarrow \omega c, \quad \quad \omega\rightarrow 0
\end{align}
\subsubsection{Carroll transformation}
Carroll transformations arise from examining the implications of reducing the speed of light $c$ to an infinitesimal value.
In order to obtain the Carroll transformations from \eqref{poincare trans3}, we use the limit (\ref{vanishing c}) and consider the scaling limit $\vec \zeta=c\vec \beta_{C}$, with the parameter $\vec \beta_{C}$ as the Carroll boost parameter. Thus we get \cite{levy1965nouvelle}:
\begin{align}
    t^{\prime}=t-\vec \beta_{C} .\vec x+a_0,\quad \vec x^{\prime}= \vec x+ \vec a.
\end{align}
Now we can derived the Carroll transformation for energy and momentum by using the limit (\ref{vanishing c}) on the Lorentz transformation on energy and momentum \eqref{poincare momen}. Thus, we obtain the following transformation rules \cite{levy1965nouvelle}:
\begin{align}
    {\vec {p}}^{\, \prime}=\vec p-\vec \beta_{C} E,\quad E^{\prime}=E
\end{align}
The general properties of these transformations are given by the following expressions:
\begin{align}
    (t,\vec x)\longmapsto (t-\vec{\beta}_{C}.\vec x+a_0,R\vec x+\vec a),
\end{align}
where the infinitesimal Carroll transformations can be written as follows \cite{Bergshoeff:2014jla,de2022carroll}:
\begin{align} \label{Carroll trans}
    \delta_Ct=-\vec \beta_{C}.\vec x+a_t,\quad \delta_C \vec x=R\vec x+\vec a,\quad \delta_C\vec p=R\vec p-\vec{\beta}_{C}E,\quad \delta_C E=0.
\end{align}
\subsubsection{Carroll limits}
Carroll particles are theoretical constructs in physics that emerge from the study of Carroll symmetry. In this context, Carroll symmetries govern the behavior of particles instead of the usual Poincaré transformations. By considering the Carrollian limit, we can identify two distinct types of Carroll particles: time-like and space-like particles. Time-like free Carroll particles do not move in the traditional sense but may have non-trivial dynamics when interacting with other particles or fields within the Carrollian framework. Space-like Carroll particles are another dynamical sector of Carroll particles that exhibit non-trivial dynamics; we will explain this sector with more details in the subsequent subsections.  \\
To derive the Carrollian particle theory, one can take the limit with the following rescalings:
\begin{align} \label{first type lim}
    x^0\rightarrow \omega t,\quad p^0 \rightarrow\frac{E}{\omega},\quad e\rightarrow+\omega^2 e,
\end{align}
to obtain the dynamics of timelike Carroll particles or take the limit with the rescalings of the spacial components:
\begin{align} \label{ second type lim}
    \vec x\rightarrow \omega \vec x,\quad \vec p \rightarrow \frac{\vec p}{\omega},\quad e\rightarrow -\omega^2 e.
\end{align}
which yeilds the dynamics of spacelike Carroll particles. We will obtain two different actions describing Carroll particles by employing the limits (\ref{first type lim}) and (\ref{ second type lim}) to the canonical action (\ref{RPPA}). 
\subsubsection{ Time-like Carroll Particles}
Time-like Carroll particles are characterized by having a non-vanishing value of energy, which is part of the broader exploration of non-Lorentzian physics. In order to be able to describe and analyze the time-like Carroll particle, one has to start from the action of the relativistic particle. 
Our initial reference is the canonical action (\ref{RPPA}) with two alternatives to rescaling the Carrollian limits (\ref{first type lim}) and (\ref{ second type lim}). If we select the scalings from (\ref{first type lim}) along with the mass rescaling $m\rightarrow\frac{M}{\omega}$, then in the limit $\omega \rightarrow 0$, we obtain the canonical action for time-like Carroll particles as follows \cite{Bergshoeff:2014jla}:
\begin{align} \label{Carroll electric action}
S=\int d\tau L_{eC}(t,E,\vec x,\vec p,e;\dot t,\dot E,\dot {\vec x},\dot {\vec p})=\int d\tau[-E\dot t+\vec {p} .\dot{\vec{x}}-\frac{e}{2}(E^2-M^2)].
\end{align}
The canonical action (\ref{Carroll electric action})  is  invariant under the Carroll transformation (\ref{Carroll trans}),
\begin{align*}
   \delta_Ct=-\vec \beta_{C}.\vec x+a_t,\quad \delta_C \vec x=R\vec x+\vec a,\quad \delta_C\vec p=R\vec p-\vec{\beta}_{C}E,\quad \delta_C E=0,
\end{align*}
while the Carroll transformation of the einbein variable is given by
\begin{align} \label{Carroll trans 2}
    \delta_C e=0.
\end{align}
Then the equation of motions of the canonical action (\ref{Carroll electric action}) are:
\begin{align} \label{time-like e.o.m}
    \dot t= -eE, \quad \dot E=0,\quad \dot {\vec x}=0,\quad \dot{\vec p}=0.
\end{align}
According to these equations of motion, these particles cannot move \cite{de2022carroll,Figueroa-OFarrill:2023vbj,Bergshoeff:2014jla}. 

The canonical Lagrangian given by equation (\ref{Carroll electric action}) remains invariant under the Carroll transformations described by (\ref{Carroll trans}) and (\ref{Carroll trans 2}). As mentioned above, the existence of a symmetry transformation implies the presence of a conservation law.
Then, the conserved current associated with the Carroll transformations (\ref{Carroll trans}) and (\ref{Carroll trans 2}), is as follows: 
\begin{align}\label{Noether 2}
    J_C^\tau=\frac{\delta L_{eC}}{\delta \dot t} \delta_Ct+ \frac{\delta L_{eC}}{\delta \dot E} \delta_CE+\frac{\delta L_{eC}}{\delta \dot {\vec x}} \delta_C \vec x+\frac{\delta L_{eC}}{\delta \dot {\vec p}} \delta_C \vec p.
\end{align}
The symmetry generators associated with the Noether current (\ref{Noether 2}) are the Carroll generators \cite{Bergshoeff:2014jla,Casalbuoni:2023bbh}:
\begin{align} \label{Carroll operators}
    \hat{H}=E,\quad \hat{P}=\vec p ,\quad \hat{B}=E\vec x , \quad \hat{J}=\vec x  \times \vec p \, .
\end{align}
If we denote the canonical momenta of the configuration variables as $\pi$, we arrive at the primary constraints,\cite{Bergshoeff:2022qkx}:
\begin{align} \label{primery 2}
    \pi_E=0,\quad \pi_t=-E,\quad \pi_{\vec{x}}=\vec p,\quad \pi_{\vec{p}}=0,\quad \pi_e=0,
\end{align}
where the Poisson brackets of variables are as follows (\ref{Carroll electric action}) \cite{Casalbuoni:2023bbh,Bergshoeff:2014jla}:
\begin{align}
    \{\ E,\ t\}=1, \quad \{\ e,\pi_e\}=1,\quad \{\ x^i, \ p^j\}=\delta^{ij} \quad i,j=1,2,3.
\end{align}
The corresponding algebra, associated with the generators in equation (\ref{Carroll operators}), is known as the Carroll algebra:
\begin{equation}\label{carroll-alg}
\begin{split}
    \{\ E,\hat{P}_{i}\} &=\{\ E,\hat{B}_{i}\} =\{\ E,\hat{J}_{ij}\}=0 , \\
      \{\hat{P}_{i}.\hat{B}_{i}\} &=-\delta_{ij}E,\; \{\hat{P}_{i},\hat{J}_{jk}\}=-\hat{P}_{k} \delta_{ij}+\hat{P}_{j}\delta_{ik},\quad \{\hat{B}_{i},\hat{J}_{jk}\}=-\hat{B}_{k}\delta_{ij}+\hat{B}_{j}\delta_{ik}, \\
      \{\hat{J}_{ij},\hat{J}_{kl}\} &=\delta_{ik}\hat{J}_{jl}-\delta_{il}\hat{J}_{jk}-\delta_{jk}\hat{J}_{il}+\delta_{jl}\hat{J}_{ik}.
      \end{split}
\end{equation}
\subsubsection{Space-like Carroll Particles}
The main feature of the space-like Carroll particles is characterized by having a vanishing energy value; now we apply the rescaling \eqref{ second type lim}, together with $m\rightarrow \frac{M}{\omega}$. After taking the limit $\omega \rightarrow 0$ of the canonical action (\ref{RPPA}), we obtain: 
\begin{align} \label{first magn Carroll action}
    S=\int\extd\tau L=\int\extd\tau[-E\dot t+\vec p.\dot{\vec x}-\frac{e}{2}(\vec{p}^{\,2}+M^2)]
\end{align}
This particular form of the action has not been previously derived using a limiting procedure. However, it has been reported in \cite{Bergshoeff:2022qkx}, where the authors employed a seed Lagrangian method to arrive at it.
It's important to note that this action is not invariant under Carroll transformations (\ref{Carroll trans}); because the mass-shell constraint $(\vec p^{\,2}+M^2)$ is not invariant under the  Carroll transformations (\ref{Carroll trans}) 
\begin{align}
\delta_C({\Vec{p}}^{\,2}+M^2)= -2{\Vec{p}}.\vec \beta E.
\end{align}
In order to make the canonical action (\ref{first magn Carroll action}) invariant with respect to the Carroll transformations (\ref{Carroll trans}) and (\ref{Carroll trans 2}), we can include a Lagrange multiplier term. Then we have
\begin{align} \label{second magn Carroll action}
    S=\int\extd\tau L_{mC}(t,E,\vec x,\vec p,e,\chi;\dot t,\dot E,\dot {\vec x},\dot {\vec p}) =\int\extd\tau[-E\dot t+\vec{p}.\dot{\vec{x}}-\frac{e}{2}({\vec{p}}^{\,2}+M^2)+\chi E]
\end{align}
where the new variable $\chi$ transforms as,
\begin{align} \label{Carroll trans 3}
    \delta_C \chi =-e\vec{p}.\vec{\beta} , 
\end{align}
under the Carroll transformations. Thus, the equations of motion are
\begin{align} \label{e.o.m 5}
    \dot E=0,\quad \dot{\vec x}=e\vec p,\quad \dot t=\chi,\quad \dot{\vec p}=0,\quad E=0.
\end{align}
According to the equations of motion, space-like Carroll particles can be in motion while possessing no energy. Then, the Noether current associated with the Carroll transformations (\ref{Carroll trans}), (\ref{Carroll trans 2}) and (\ref{Carroll trans 3}) is as follows:
\begin{align}
    J_C^\tau=\frac{\delta L_{mC}}{\delta \dot t} \delta_Ct+ \frac{\delta L_{mC}}{\delta \dot E} \delta_CE+\frac{\delta L_{mC}}{\delta \dot {\vec x}} \delta_C \vec x+\frac{\delta L_{mC}}{\delta \dot {\vec p}} \delta_C \vec p.
\end{align}
The symmetry generators are the same as the time-like limit symmetry generators in (\ref{Carroll operators}). Therefore, we have:
\begin{align} 
    \hat{H}=E,\quad \hat{P}=\vec p ,\quad \hat{B}=E\vec x , \quad \hat{J}=\vec x  \times \vec p,
\end{align}
with the primary constraints:
\begin{align}\label{primary 3}
    \pi_E=0,\quad \pi_t=-E,\quad \pi_{\vec{x}}=\vec p,\quad \pi_{\vec{p}}=0,\quad \pi_e=0,\quad \pi_\chi=0.
\end{align}
This produces the same Carroll algebra \eqref{carroll-alg} in this space-like limit.
\section{Non-Lorentzian particles in Electromagnetic field}\label{Non-Relativistic point particle in EM}
In this section, we extend the theory of non-relativistic charged particles to include their interaction with electromagnetic fields. We will begin with a relativistic charged particle in an electromagnetic field and examine its limiting cases to uncover the fracton/non-relativistic particle dualities in the presence of gauge fields. 
\subsection{Relativistic charged particle}
We speculate that some of the charged quasi-particles, subsystems that effectively behave as charged particles, might display Carrollian or Galilean dynamics across various energy regimes. To study these systems, we start from the relativistic point particle described by the canonical action (\ref{RPPA}) having charge $q$ which is under the influence of the electromagnetic field.
In this case, our interaction term, which is Poincaré invariant, is as follows:
\begin{align}
    L_{int}=q A_{\mu}.U^\mu,
\end{align}
where the four-vector $U^\mu$ is defined as the four-velocity of our point particle that follows as:
\begin{align}\label{four-velocity}
    U^\mu=\frac{d x^\mu}{d\tau}=\dot x^{\mu}.
\end{align}
Now, the canonical action of the relativistic charged point particles is \cite{Deriglazov:2017biu}
\begin{align} \label{Electromagnetic action}
    S=\int\extd \tau [-p_\mu \dot x^{\mu}-\frac{e}{2}(p^2-m^2)+q A_{\mu} U^\mu].
\end{align}
We can rewrite the canonical action (\ref{Electromagnetic action}) by utilizing equation (\ref{four-velocity}) and defining the new momentum variable $p_{EM}^\mu=p^\mu-qA^\mu$ as the momentum of a charged point particle. Then, the canonical action (\ref{Electromagnetic action}) can be re-expressed in a new form as follows \cite{Deriglazov:2017biu}:
\begin{align}\label{Electromagnetic action 2}
    S=\int\extd \tau [-(p_\mu-qA_\mu) \dot x^{\mu}-\frac{e}{2}((p^\mu-qA^\mu)^2-m^2)].
\end{align}
The equation of motions of the action (\ref{Electromagnetic action}) are
\begin{equation}
    \dot x^\alpha =- ep^\alpha,\quad \dot p_\alpha=qU^{\mu}F_{\mu\alpha} 
\end{equation}
where we have $\dot{p}_{0}=q\, \gamma \, \vec{u}\cdot\vec{E}$ and
\begin{align*}
    A^{\mu}=\begin{pmatrix}
    \frac{\phi}{c}\\\vec A
\end{pmatrix} ; \quad     U^{\mu}=\begin{pmatrix}
    \gamma_{\zeta}c\\ \gamma_{\zeta}\vec u
\end{pmatrix}; \quad     \dot{A^{\mu}}=U^{\alpha}\frac{\partial A^{\mu}}{\partial x^{\alpha}} \, .
\end{align*}
The Lorentz force law, which describes the force on a charged particle due to electric and magnetic fields, is given by: $\dot p^{\mu}=q F^{\mu\nu}U_{\nu}$, in which $F^{\mu\nu}$ is the electromagnetic tensor. We can rewrite this equation of motion in terms of the electric and magnetic fields,
\begin{align}
     \dot {\vec p}=q\gamma_{\zeta} (\vec E+\vec u\times \vec B),
\end{align}
where $\vec u$ is the velocity of the relativistic particles. 
\subsection{Charged Carroll Particle}
Carrollian electrodynamics is a branch of theoretical physics that studies the behavior of electromagnetic fields and charged particles in the context of Carroll symmetry, where the speed of light is effectively zero. The behavior of electric and magnetic fields in Carroll symmetry can be quite different. For example, in Maxwell's theory, there are two different Carroll limits at the level of action, corresponding to the electric and magnetic limits. The question that remains is whether there exists an Electromagnetism theory that is invariant under Carroll symmetry.
In their study \cite{Duval:2014uoa}, it was shown that there is an electromagnetic theory that respects Carroll symmetry, known as ``Carroll Electromagnetism''.

\subsubsection{Electric Charged Carroll Particle}
By applying the limit (\ref{first type lim}) and rescaling
\begin{align}
    \phi\rightarrow \frac{1}{\omega} \phi, 
\end{align}
in the canonical action (\ref{RPPA}),
we find a well-defined limit that straightforwardly yields Carroll-invariant theory. We refer to this theory as the time-like Carroll particle with no charge. Now, we follow the same procedure for the charged particles to derive the corresponding Carroll invariant theory. This theory describes point particles with electric charge, which we refer to \textit{Electric Carroll particles}.
The canonical action of the Electric Carroll particle can be derived as the following,
\begin{align}\label{action-Ecarroll}
    S_{EC}=\int\extd \tau[-E\dot t+\vec p.\dot{\vec x}-\frac{e}{2}(E^2-M^2)+q\dot{t}\phi(\vec x,t)-q\dot{\vec x}\cdot\vec A(\vec x,t)] \, ,
\end{align}
which is invariant under the Carroll symmetry and the equations of motion are
\begin{equation}\label{e.o.m:electric-carroll}
\begin{split}
    \dot t &=eE, \quad \dot{\vec x}=0,\quad \dot E=q\dot \phi(\vec x)-q\dot{t}\partial_{t}\phi(\vec x,t)+q\dot{\vec x}.\partial_{t}\vec A=0,\\
        \dot{ p}_{m} &=q\dot{A}_{m}(\vec x,t)+q\dot{t}\partial_{m}\phi(\vec x,t)-q\dot{x}^{i}\partial_{m}A_{i}.
        \end{split}
\end{equation}
The Loretnz force in this case can be rewritten as the following
\begin{align}\label{ECCP:LF}
        \dot{p}_{m}=q\dot{x}^{i}(\partial_{i}A_{m}-\partial_{m}A_{i})+q\dot{t}(\partial_{m}\phi+\partial_{t}A_{m})
\end{align}
Equations of motion (\ref{e.o.m:electric-carroll}) demonstrate that the Lorentz force acting on Electric Carroll particles is exclusively electrostatic when the gauge $t = \tau$ is applied. Magnetic fields have no influence on Electric Carroll particles, as these particles are immobile. Consequently, with the following definition of the electric field:
\begin{align}
    \vec{E}=\frac{\partial \vec A}{\partial t}+\vec{\nabla}\phi,
\end{align}
the Lorentz force \eqref{ECCP:LF} will be
\begin{align}
    \dot{\vec p}=q\vec{E} .
\end{align}
The interesting result is that the electric Carroll particles are not able to move in the electromagnetic field and the energy $E$ (Mass in some terminology) is conserved. These results were previously reported in \cite{Marsot:2022imf,Marsot:2021tvq} but derived from a different procedure.  
The canonical action (\ref{action-Ecarroll}) is invariant under the Carroll transformation (\ref{Carroll trans}) where we have:
\begin{align*}
   \delta_C t=-\vec \beta_{C} \cdot\vec x + a_t, \quad \delta_C \vec x= \vec a, \quad\delta_C \vec p=-\vec \beta_{C} E, \quad\delta_C E=0,
\end{align*}
and for the electromagnetic fields, we find the following transformation rules,
\begin{align}\label{Carroll trans 5}
    \delta_C \vec A=-\vec \beta_{C} \phi, \quad\delta_C \phi=0.
\end{align}
Then the Noether current is:
\begin{align}\label{Noether 7}
    J_C^\tau=\frac{\delta L_{EC}}{\delta \dot t} \delta_Ct+ \frac{\delta L_{EC}}{\delta \dot E} \delta_CE+\frac{\delta L_{EC}}{\delta \dot {\vec x}} \delta_C \vec x+\frac{\delta L_{EC}}{\delta \dot {\vec p}} \delta_C \vec p \, ,
\end{align}
with the following associated symmetry generators that form the Carroll-Maxwell generators:
\begin{align} \label{Carroll operators 2}
    \hat{H}=E-q\phi,\quad \hat{P}=\vec p-q\vec A ,\quad \hat{B}=(E-q\phi)\vec x , \quad \hat{J}=\vec x  \times (\vec p-q\vec A).
\end{align}
\subsubsection{Magnetic Charged Carroll Particle}
In the previous subsection, we identified the \textit{Electric Carroll} particles by applying the first type of Carroll limit to the canonical action (\ref{Electromagnetic action 2}). Now, by applying the second type of Carroll limits \eqref{ second type lim} and rescaling $\vec A\rightarrow \frac{\vec A}{\omega}$, we find the action
\begin{align}\label{action:MCC1}
    S_{MC}=\int\extd \tau[-E\dot{t}+\vec p.\dot{\vec x}-\frac{e}{2}(\vec p^{\ 2}+M^2)+q\dot t\phi(\vec x,t)-q\dot{\vec x} \cdot\vec A(\vec x.t)].
\end{align}
However, as with the same action (\ref{first magn Carroll action}), this action is not invariant under the Carroll transformations (\ref{Carroll trans}) and (\ref{Carroll trans 5}) because the mass-shell constraint $ (\vec p^{\ 2}+M^2)$ is not invariant under the Carroll transformations.
We can include a Lagrange multiplier term to ensure that the canonical action (\ref{action:MCC1}) is invariant under the Carroll transformations. Then, we propose the following action for the \textit{Magnetic Carroll} particles,
\begin{align}\label{Magnetic Carroll particles}
    S_{MC}=\int\extd \tau[-E\dot{t}+\vec p.\dot{\vec x}-\frac{e}{2}(\vec p^{\ 2}+M^2)+q\dot t\phi(\vec x,t)-q\dot{\vec x}\cdot\vec A(\vec x,t)+\chi (E-\theta)],
\end{align}
which is now invariant due to the Lagrange multiplier $\chi$. Thus, the equations of motion are 
\begin{equation}\label{E.O.Ms Magnetic Carroll particles}
\begin{split}
    \dot t &=\chi, \quad \dot{\vec x}=e\vec p,\quad \dot E=q\dot \phi(\vec x)-q\dot{t}\partial_{t}\phi(\vec x,t)+q\dot{\vec x}.\partial_{t}\vec A =q\dot{\vec x}(\vec{\nabla}\phi+\partial_{t}\vec{A})=q\dot{\vec x}\cdot\vec E ,\\
        \dot{ p}_{m} &=q\dot{A}_{m}(\vec x,t)+q\dot{t}\partial_{m}\phi(\vec x,t)-q\dot{x}^{i}\partial_{m}A_{i} \, ,
        \end{split}
\end{equation}
where $\chi$ behaves under the Carroll transformations as follows:
\begin{align}
    \delta_C \chi=-e\vec \beta_{C}\vec p .
\end{align}
If we apply the gauge fixing $t=\tau$, the Lorentz force on the Magnetic Carroll particles is,
\begin{align*}
    \dot{p}_{m}&=q(\dot{t}{\partial_{t}A_{m}}+\dot{x}^{i}\partial_{i}A_{m})+q\dot{t}\partial_{m}\phi-q\dot{x}^{i}\partial_{m}A_{i}\\
    &=q(\vec{E}+\dot{\vec{x}}\times \vec{B}) \, .
    \end{align*}
In this case, unlike Electric Carroll particles, Magnetic Carroll particles are under the influence of the magnetic field, which means the Lorentz force acting on magnetic Carroll particles includes electromagnetic forces.
Magnetic sector of charged Carroll particles has not been reported before. Important points can be observed from these equations; first, particles can move, energy is not conserved, and particles experience accelerations induced by electric and magnetic fields.
Within the theoretical framework of Carroll symmetry, magnetic Carroll particles can exhibit motion under certain conditions. The motion of Magnetic Carroll particles is especially interesting in the context of black hole physics, in which the geometry of the horizon can be described using Carrollian geometry \cite{Bicak:2023rsz}.
\subsection{Charged Galilean Particles}
In the context of electromagnetism, Galilean electromagnetism attempts to describe electric and magnetic fields in situations where charged bodies move at nonrelativistic velocities, that is, velocities much slower than light \cite{levy1971galilei,LeBellac:1973unm}. There exists a non-relativistic electromagnetic theory called ``Galilean Electromagnetism'' \cite{LeBellac:1973unm}. In general, Galilean electromagnetism is a formal electromagnetic theory that is consistent with Galilean invariance. Galilean electromagnetism is useful for describing the electric and magnetic fields in the vicinity of charged bodies moving at non-relativistic speeds relative to the frame of reference. The resulting mathematical equations are simpler than the fully relativistic forms because certain coupling terms are neglected \cite{rousseaux2013forty}.  
\subsubsection{Magnetic Charged Galilean Particles}
 
In order to describe the dynamics of Galilean charged particles in electromagnetic fields, we consider both the Galilean limits (\ref{Galelian first type lim}) and (\ref{Galelian second type lim}) for the canonical action (\ref{Electromagnetic action 2}).
For Magnetic Galilean particles, we apply the Galilean limit (\ref{Galelian first type lim}) and rescaling $m \rightarrow \frac{\kappa}{\omega}$, thus we obtain the following invariant action
\begin{align}\label{Galilean magnetic action}
    S_{MG}=\int\extd \tau[-E\dot{t}+\vec{p}\cdot\dot{\vec x}-\frac{e}{2} (\vec {p}^{\ 2} + \kappa^2)+q \, \dot t\phi(\vec x,t)-q \, \dot{\vec x}\cdot\vec A(\vec x,t)] ,
\end{align}
with the following equations of motion,
\begin{equation}\label{e.o.m:MGalilean}
\begin{split}
    \dot t &=0,\quad \dot{\vec x}=e\vec p, \\
    \dot E &=q\dot \phi(\vec{x},t)-q\dot{t}\partial_{t}\phi+q\dot{\vec x}\cdot\partial_{t}\vec A=q\dot{\vec x}\cdot\vec{E} , \\
        \dot{ p}_{m} & =q\dot{A}_{m}(\vec x,t)+q\dot{t}\partial_{m}\phi(\vec x,t)-q\dot{x}^{i}\partial_{m}A_{i}. 
        \end{split}
\end{equation}
According to equations (\ref{e.o.m:MGalilean}), the Lorentz force acting on Magnetic Galilean particles can be re-expressed as:
\begin{align*}
        \dot{p}_{m}&=q\dot{x}^{i}\partial_{i}A_{m}-q\dot{x}^{i}\partial_{m}A_{i}\\
    &=q\dot{x}^{i}(\partial_{i}A_{m}-\partial_{m}A_{i})\\
    &=q\dot{\vec{x}}\times \vec{B},
    \end{align*}
This result aligns with expectations based on the equations of motion, demonstrating that Magnetic Galilean particles are transparent to electric fields.
We see that the Magnetic Galilean massless particles are able to move in the electromagnetic field. The canonical action (\ref{Galilean magnetic action}) is invariant under the Galilean transformation (\ref{Galilean trans}), as follows
\begin{align} 
    \delta_Gt=a_t,\quad \delta_G \vec x=R\vec x+\vec \beta_{G} t+\vec a,\quad \delta_G\vec p=R\vec p,\quad \delta_G E=\vec \beta_{G}.\vec p,
\end{align}
where we find the following transformation rules for electromagnetic potentials,
\begin{align} \label{Galilean trans 4}
    \delta_{G}\vec A=0,\quad \delta_{G}\phi=-\vec \beta_{G} \cdot \vec A,
\end{align}
These relations should be compared with equations \eqref{Carroll trans 5}, where instead of a vector potential, the scalar potential changes under Galilean transformations. 
The Noether current is
\begin{align} \label{Noether 8}
    J_C^\tau=\frac{\delta L_{MG}}{\delta \dot t} \delta_{G}t+ \frac{\delta L_{MG}}{\delta \dot E} \delta_{G}E+\frac{\delta L_{MG}}{\delta \dot {\vec x}} \delta_{G} \vec x+\frac{\delta L_{MG}}{\delta \dot {\vec p}} \delta_{G} \vec p \, ,
\end{align}
with the Galilean-Maxwell generators:
\begin{align} \label{Galilean operators 2}
    \hat{H}=E-q\phi,\quad \hat{P}=\vec p-q\vec A ,\quad \hat{B}=(\vec p-q\vec A)t , \quad \hat{J}=\vec x  \times (\vec p-q\vec A).
\end{align}
\subsubsection{Electric Charged Galilean Particles}
Electric Galilean particles are charged particles can be obtained in the framework of Galilean electrodynamics.
If we apply the second type of Galilean limit, Eq.~\eqref{Galelian second type lim}, along with the rescaling $m \rightarrow \frac{\epsilon}{\omega}$ to the relativistic action, we obtain:
\begin{align}
     t\rightarrow \frac{ t}{\omega},\quad E \rightarrow {\omega}{E},\quad \phi\rightarrow\omega \phi,\quad e\rightarrow -\frac{ e}{\omega^2},\quad\omega\rightarrow\infty ,
\end{align}
we will find the following action
\begin{align} \label{Galilean Electric action}
    S_{EC}=\int\extd \tau[-E\dot{t}+\vec p\cdot \dot{\vec x}-\frac{e}{2} (E^2 - \epsilon^2)+q\dot t\phi(\vec x,t)-q\dot{\vec x}\cdot\vec A(\vec x,t) ] \, .
\end{align}
This action is not invariant under the Galilean transformations (\ref{Galilean trans}) and (\ref{Galilean trans 4}), because the term $(E^2 - \epsilon^2)$ is not invariant under the Galilean transformations (\ref{Galilean trans}) and (\ref{Galilean trans 4}). To make the theory invariant under the Galilean transformations, as observed in Magnetic Carroll particles, we can add a Lagrange multiplier term to ensure that the canonical action (\ref{Galilean Electric action}) remains invariant under the Galilean transformations. Then we construct the following action for the theory,
\begin{align} \label{Galilean Electric action 1}
    S_{EC}=\int\extd \tau[-E\dot{t}+\vec p\cdot \dot{\vec x}-\frac{e}{2}(E^2 - \epsilon^2)+q\dot t\phi(\vec x,t)-q\dot{\vec x}\cdot\vec A(\vec x,t) +\vec \chi \cdot \vec p],
\end{align}
while the equations of motion are 
\begin{equation} \label{e.o.ms of Galilean Electric action 1}
\begin{split}
    \dot{t} &=eE,\quad \dot {\vec x}= \vec \chi , \\
    \dot E &=q\dot \phi(\vec{x},t)-q\dot{t}\partial_{t}\phi+q\dot{\vec x}\cdot\partial_{t}\vec A=q\dot{\vec x}\cdot\vec{E} , \\
        \dot{ p}_{m} &=q\dot{A}_{m}(\vec x,t)+q\dot{t}\partial_{m}\phi(\vec x,t)-q\dot{x}^{i}\partial_{m}A_{i}.
        \end{split}
    \end{equation}
By gauge fixing $t=\tau$, we find  
\begin{align*}
        \dot{p}_{m}&=q(\dot{t}{\partial_{t}A_{m}}+\dot{x}^{i}\partial_{i}A_{m})-q\dot{t}\vec{\nabla}\phi-q\dot{x}^{i}\partial_{m}A_{i}\\
    &=q(\vec{E}+{\vec{\chi}}\times \vec{B}),
    \end{align*}
where the transformation rule for the new variable $\chi$ is 
\begin{align}
    \delta_{G} \chi=-e\vec \beta_{G}E.
\end{align}
In contrast to Electric Carroll particles, Electric Galilean particles experience both electric and magnetic fields in the Lorentz force. This distinction arises from the unique characteristics of each particle type.
\subsection{Fracton/Non-Lorentzian Duality}\label{FNLduality}
In this section, we explore a key motivation underlying our work: the construction of a duality between fractons and non-Lorentzian particles coupled to gauge fields. A particularly intriguing feature of fracton behavior is their dual relationship with other theoretical particle types. Extending this duality to gauge-coupled scenarios significantly enhances our understanding of these emergent phenomena.
\subsubsection{Fracton/Carroll Duality}

There is a concept called the Carroll/Fracton duality \cite{Figueroa-OFarrill:2023vbj}, which studies the relationship between the Carroll and Fracton symmetries. This duality has implications for the classification of different types of fracton particles and can provide insights into their properties. For example, it explains how the fracton electric charge and dipole moment correspond to specific energy and center-of-mass parameters in the Carroll framework \cite{Figueroa-OFarrill:2023vbj}. 

\paragraph{Isolated Fracton/Time-like Carroll Particles:} The action for free fractons, given by equation \eqref{Isolated Fractons}, and their equations of motion (equation \eqref{E.O.Ms of Isolated Fractons}), exhibit dipole symmetry. We can compare these with the action and equations of motion for Time-Like Carroll particles.  The canonical action for Carroll particles is presented in equation (\ref{Carroll electric action}), and their equations of motion are shown in \eqref{time-like e.o.m}.
This comparison reveals a duality between isolated fractons and Time-Like Carroll particles at the dynamical level, which shows these distinct particles are fundamentally immobile, meaning they cannot move in the traditional sense.\footnote{If we define modified momentum $\widetilde p$ in isolated fractons, the duality can be visualized.}

\paragraph{Static Dipole/Electric Charged Carroll Particles:} A static dipole, as described by equation \eqref{Static Dipole} and its corresponding equations of motion \eqref{fracton e.o.m 1}, can be compared to Electric Charged Carroll particles, which are defined by the canonical action \eqref{action-Ecarroll} and their equations of motion \eqref{e.o.m:electric-carroll}. Both types of particle are immobile and subject to the Lorentz force, which in this case consists solely of the electrostatic component. 
Unlike the study in \cite{Figueroa-OFarrill:2023vbj} of free fractons, we find that Electric Charged Carroll particles remain immobile even when exposed to electromagnetic fields.

\paragraph{Mobile Dipole/Magnetic Charged Carroll Particles:} We can compare the second model of mobile dipoles, described by equation \eqref{Mobile Dipole 2} and its corresponding equations of motion \eqref{E.O.Ms Mobile Dipole 2}, with Magnetic Charged Carroll particles defined by the action \eqref{Magnetic Carroll particles} and their equations of motion \eqref{E.O.Ms Magnetic Carroll particles}. Both types of particles show similarity at the dynamical level and exhibit mobility, and the Lorentz force acting on them includes both electric and magnetic components. This comparison highlights a duality between these two models, distinguishing them from the previously discussed immobile particles.

\subsubsection{Fracton/Galilean Duality}

One of the intriguing dualities that can exist is the relationship between fracton symmetry and Galilean symmetry, which we refer to as the Galilean/Fracton duality. The Carroll group and the Galilean group are isomorphic in three dimensions ($D=3$); thus, this isomorphism implies that fracton symmetry possesses a duality with the Galilean group in $D=3$ \cite{Vadali:2023raq,Spieler:2023wkz,Zhang:2023eaw}. This leads to an important question: Which fracton particles or fracton excitations are dual to the Galilean particles?
We will clarify in the following that only mobile dipoles can be dual to Galilean particles. 

\paragraph{Mobile Dipole/Magnetic Charged Galilean Particles:} The preliminary model of mobile dipoles, as described in \eqref{Mobile Dipole 1} along with its corresponding equations of motion \eqref{E.O.Ms Mobile Dipole 1}, can also be closely compared to Magnetic Charged Galilean particles, which are defined by \eqref{Galilean magnetic action} and their equations of motion \eqref{e.o.m:MGalilean}. Both types of particles are mobile, and the Lorentz force accounts for the magnetic force component.
\paragraph{Mobile Dipole/Electric Charged Galilean Particles:} A second model of mobile dipoles, as described in \eqref{Mobile Dipole 2}, together with its corresponding equations of motion \eqref{E.O.Ms Mobile Dipole 2}, can be closely compared to Electric Charged Galilean particles, which are defined by \eqref{Galilean Electric action 1} and their equations of motion \eqref{e.o.ms of Galilean Electric action 1}. Both types of particles are mobile, and the Lorentz force accounts for both electric and magnetic force components. The interesting aspect of this comparison is that Electric Charged Galilean and Magnetic Charged Carroll particles are also dual to each other \cite{Gomis:2022spp}.


\subsubsection{Lineon/Non-Lorentzian Duality}
Lineon particles can move freely along a one-dimensional line. In this subsection, we will address the duality between lineons and non-lorentzian particles.
\paragraph{Lineon/Magnetic Charged Carroll and Electric Charged Galilean Particles:} Lineon particles, as previously mentioned, can be described by the canonical action \eqref{action:lineon} and their corresponding equations of motion \eqref{e.o.ms of lineons}. By conserving the angular and linear momentum of the vector charge, we find that lineons can only move along their own vector charge $\vec{q}$. Lineons can be closely compared to Magnetic Charged Carroll and Electric Charged Galilean particles \eqref{E.O.Ms Magnetic Carroll particles} and \eqref{Galilean Electric action 1} respectively. In this scenario, where particles move in the $\vec{\chi}$ direction, a consistent theory requires the electric field to be perpendicular to the direction of motion. Consequently, the Lorentz force acts perpendicularly to the motion of the particles.

\section{Fracton and Non-Lorentzian Algebras}\label{algebra}
This section has two primary goals: first, to investigate the algebraic structure of fractons and their connections to non-Lorentzian algebras; and second, we extend the fracton algebra to incorporate the Maxwell algebra, establishing a theoretical foundation for the preceding sections concerning the coupling of fractons with gauge fields.

\subsection{Carrollean and Galilean algebras}
Non-Lorentzian algebras are mathematical structures that arise when studying physical theories with symmetries different from those of Lorentzians. They are associated with kinematic spacetimes that do not obey the usual Lorentz symmetry, such as the Galilean and Carrollian spacetimes. These algebras play a crucial role in understanding the kinematics and dynamics of non-Lorentzian theories, which are applied in areas such as condensed matter physics, holography and gravitational waves \cite{Bergshoeff:2022eog,Lambert:2021nol}. In the following, we will briefly mention these two non-Lorentzian algebras.
\paragraph{Carrollean algebras} has the following structure \cite{Bergshoeff:2022eog,Figueroa-OFarrill:2022nui,Figueroa-OFarrill:2018ilb}:
\begin{equation}\label{carroll-algebra}
\begin{split}
    \{\ E,\hat{P}_{i}\} &=\{\ E,\hat{B}_{i}\} =\{\ E,\hat{J}_{ij}\}=0 , \\
      \{\hat{P}_{i}.\hat{B}_{i}\} &=-\delta_{ij}E,\; \{\hat{P}_{i},\hat{J}_{jk}\}=-\hat{P}_{k} \delta_{ij}+\hat{P}_{j}\delta_{ik},\quad \{\hat{B}_{i},\hat{J}_{jk}\}=-\hat{B}_{k}\delta_{ij}+\hat{B}_{j}\delta_{ik}, \\
      \{\hat{J}_{ij},\hat{J}_{kl}\} &=\delta_{ik}\hat{J}_{jl}-\delta_{il}\hat{J}_{jk}-\delta_{jk}\hat{J}_{il}+\delta_{jl}\hat{J}_{ik}.
      \end{split}
\end{equation}
which includes generators for spatial translations, time translations, rotations, and Carrollian boosts. 
\paragraph{Galilean algebras} are given as \cite{levy1965nouvelle,Duval:2014uoa,Bagchi:2017yvj,Figueroa-OFarrill:2022nui,Bergshoeff:2022eog,Figueroa-OFarrill:2024ocf}:
\begin{equation}\label{galilean-algebra}
\begin{split}
    \{\ E,\hat{P}_{i}\} & =\{\ E,\hat{J}_{ij}\}= \{\hat{P}_{i}.\hat{B}_{j}\}=0 , \\
      \ \{\ E,\hat{B}_{i}\} &= \hat{P}_{i},\quad
    \{\hat{P}_{i},\hat{J}_{jk}\}=-\hat{P_k} \delta_{ij}+\hat{P}_{j}\delta_{ik},\quad \{\hat{B}_{i},\hat{J}_{jk}\}=-\hat{B}_{k}\delta_{ij}+\hat{B}_{j}\delta_{ik}, \\
    \{\hat{J}_{ij},\hat{J}_{kl}\} &=\delta_{ik}\hat{J}_{jl}-\delta_{il}\hat{J}_{jk}-\delta_{jk}\hat{J}_{il}+\delta_{jl}\hat{J}_{ik}\, .
    \end{split}
\end{equation}
These two algebras are isomorphic in $D = 3$ (where $D$ is the dimension of spacetime) \cite{Figueroa-OFarrill:2022nui,Trzesniewski:2023vou}. The isomorphism in lower dimensions is particularly interesting because it suggests a duality between the ultra-relativistic limit (Carroll) and the non-relativistic limit (Galilei) of spacetime symmetries.

\subsection{Non-Lorentzian Maxwell algebras}
Relativistic particles coupled to an electromagnetic field enjoy symmetries that extend beyond the usual Poincaré algebra symmetries \cite{Bacry:1970ye,Schrader:1972zd}. Any relativistic theory must be symmetric under the transformations of the Poincaré group, as it is the spacetime symmetry group of special relativity. So, if we want to develop a relativistic theory with greater symmetry, as mentioned in the introduction, we need to extend the Poincaré group and, consequently, its algebra. The Maxwell algebra is such an extension.

\paragraph{Maxwell algebra:}Poincaré algebra is defined in (\ref{Poincare algebra}),
assuming transformation rules for electromagnetic fields, one obtains a non-central extension of the Poincaré algebra by an antisymmetric tensor generator $\hat{Z}_{\mu \nu}$ that transforms covariantly under the Lorentz group. The resulting algebra has been referred to as the Maxwell algebra that follows \cite{Gomis:2017cmt}
\begin{equation}\label{Maxwell Algebra}
\begin{split}
    \{\hat{P_\mu},\hat{P_\nu}  \} &=\hat{Z}_{\mu \nu},\quad \{\hat{P_\mu},\hat{Z}_{\nu \rho}  \}=0,\quad \{\hat{Z}_{\mu \nu },\hat{Z}_{\rho \sigma}\}=0 , \\
    \{\hat{M}_{\mu \nu},\hat{Z}_{\rho \sigma}\} &=\eta_{\nu \rho}\hat{Z}_{\mu \sigma}+\eta_{\mu \sigma}\hat{Z}_{\nu \rho}-\eta_{\mu \rho}\hat{Z}_{\nu \sigma}-\eta_{\nu \sigma}\hat{Z}_{\mu \rho}.
    \end{split}
\end{equation}
Here, $\eta_{\mu\nu}$ is the flat Minkowski metric, and $\hat{M}_{\mu\nu} $ denote the Lorentz generators.
The Maxwell algebra represents a closed algebra and a tensorial extension of the Poincaré algebra \cite{Gomis:2017cmt}. In this particular algebra, we have $\hat{Z}_{\mu\nu}=\partial_{\mu}A_{\nu}-\partial_{\mu}A_{\nu}$.
\paragraph{Galilei-Maxwell algebra:} 
A non-relativistic Maxwell-type extension emerges in an electromagnetic background which is called the ``Galilei-Maxwell'' algebra \cite{Beckers:1983gp,Bacry:1970du,Barducci:2019fjc,Gomis:2019fdh,Cerdeira:2023ztm}. The Galilei-Maxwell algebra, obtained by contracting the relativistic Maxwell algebra (\ref{Maxwell Algebra}), follows as \cite{Beckers:1983gp,Bacry:1970du,Barducci:2019fjc,Gomis:2019fdh,Cerdeira:2023ztm}: 
\begin{equation} \label{Galilei-Maxwell}
\begin{split}
    \{\hat{P}_{i},\hat{P_j}\} &=\hat{Z}_{ij},\quad    \{\hat{H},\hat{P
    }_{i}\}=-\hat{Z}_{i},\quad\{\hat{H},\hat{B}_{i}\}=\hat{P}_{i} ,\\
    \{\hat{P}_{i},\hat{J}_{jk}\} &=-\hat{P}_{k} \delta_{ij}+\hat{P}_{j}\delta_{ik},\quad  \{\hat{B}_{i},\hat{J}_{jk}\}=-\hat{B}_{k}\delta_{ij}+\hat{B}_{j}\delta_{ik} , \\
    \{\hat{J}_{ij},\hat{J}_{kl}\}&=\delta_{ik}\hat{J}_{jl}-\delta_{il}\hat{J}_{jk}-\delta_{jk}\hat{J}_{il}+\delta_{jl}\hat{J_{ik}} , \\
    \{\hat{J}_{ij},\hat{Z}_{kl}\} &=\delta_{ik}\hat{Z}_{jl}-\delta_{il}\hat{Z}_{jk}-\delta_{jk}\hat{Z}_{il}+\delta_{jl}\hat{Z}_{ik}.
    \end{split}
\end{equation}
In Galilei-Maxwell algebra, we have new generators defined as \( \hat{Z}_{ij} = \partial_{i} A_{j} - \partial_{j} A_{i} \) and \( \hat{Z}_{i} = \partial_{i} \phi + \partial_{t} A_{i} \). 
This algebra includes the usual generators of the Galilean algebra - space translations, time translations, rotations, and Galilean boosts - along with additional generators that consider the electromagnetic field. These additional generators extend symmetries by being invariant under electromagnetic gauge transformations.\\
The Galilei-Maxwell algebra is an algebraic structure that is an extension of the Galilei algebra with respect to the principles of electromagnetism under non-relativistic conditions. The Galilei-Maxwell algebra allows us to explore how electromagnetic phenomena would behave in a theoretical framework where the speed of light is not the ultimate speed limit and where the effects of special relativity can be neglected. This is particularly useful in condensed matter physics and other areas where the velocities involved are much smaller than the speed of light and the relativistic effects are minimal. In summary, the Galilei-Maxwell algebra provides a way to understand electromagnetism in a non-relativistic context, which gives a theoretical basis for the study of systems where relativistic effects are not significant \cite{Gomis:2019fdh,McGreevy:2022oyu,goldin2001galilean}.

\paragraph{Carroll-Maxwell algebra:} An ultra-relativistic Maxwell-type extension in an electromagnetic background introduced in \cite{Barducci:2019fjc,Concha:2021jnn}, which is referred to as the ``Carroll-Maxwell'' algebra \footnote{Here, we derive the $k = 1$ contraction of the Maxwell algebra. For the $k$ contraction related to the Carroll case of the Maxwell algebra, refer to \cite{Barducci:2019fjc}.}:
\begin{equation} \label{Carroll-Maxwell}
\begin{split}
    \{\hat{P}_{i},\hat{P}_{j}\} &= \hat{Z}_{ij},\quad    \{\hat{H},\hat{P}_{i}\}=-\hat{Z}_{i},\quad\{\hat{P_i},\hat{B_j}\}=-\hat{H}\delta_{ij} , \\
    \{\hat{P}_{i},\hat{J}_{jk}\} &=-\hat{P}_{k} \delta_{ij}+\hat{P}_{j}\delta_{ik},\quad  \{\hat{B}_{i},\hat{J}_{jk}\}=-\hat{B}_{k}\delta_{ij}+\hat{B}_{j}\delta_{ik} , \\
   \{\hat{J}_{ij},\hat{J}_{kl}\}&=\delta_{ik}\hat{J}_{jl}-\delta_{il}\hat{J}_{jk}-\delta_{jk}\hat{J}_{il}+\delta_{jl}\hat{J_{ik}} , \\
    \{\hat{J}_{ij},\hat{Z}_{kl}\} &=\delta_{ik}\hat{Z}_{jl}-\delta_{il}\hat{Z}_{jk}-\delta_{jk}\hat{Z}_{il}+\delta_{jl}\hat{Z}_{ik}.
    \end{split}
\end{equation}
In order to incorporate electromagnetic fields within the ultra-relativistic framework, additional generators are introduced into the Carroll-Maxwell algebra \cite{Concha:2021jnn}, which are \( \hat{Z}_{ij} = \partial_{i} A_{j} - \partial_{j} A_{i} \) and \( \hat{Z}_{i} = \partial_{i} \phi + \partial_{t} A_{i} \). \\
The Carroll-Maxwell algebra is an intriguing extension of the Carroll algebra that integrates electromagnetism into the ultra-relativistic framework, opposite to the known non-relativistic (Galilean) one. The Carroll-Maxwell algebra helps to understand how Maxwell's equations of electromagnetism might be modified in a Carrolllian spacetime. It provides insight into the symmetries of these equations in different spacetime structures \cite{Concha:2021jnn}.
\subsection{Fracton Algebra}
%
The algebraic structure of fractonic systems is related to their unique symmetries and conservation laws. Fractons are associated with a different set of symmetries that constrain their dynamics. The algebraic framework for fractions often involves tensor gauge theories and spatial symmetries that are more complex than those found in traditional field theories. Before introducing its Maxwell-type extension, we briefly mention fracton algebra and its structure. 
In the previous section, as mentioned earlier, the fractions adhere to charge and dipole conservation laws, with $\hat{Q}$ and $\hat{Q}_{i}$ serving as symmetry generators for these conservation laws. In addition to these two conservation laws, we assume that the system is translationally invariant, which implies the conservation of momentum. Furthermore, there are also invariant under rotations and translations in space and time. This symmetry group is referred to as the ``fracton'' symmetry, which has the following commutation relations \cite{Armas:2023ouk}
\begin{equation} \label{Fracton-alg}
  \begin{split} 
    \{\hat{Q},\hat{Q_i}\} &= \{\hat{Q},\hat{P}_{i}\}=0,\quad \{\hat{P}_{i},\hat{Q}_{j}\}=\delta_{ij}\hat{Q} , \\
     \{\hat{P}_{i},\hat{J}_{jk}\} &=\delta_{ik}\hat{P}_{j}-\delta_{ij}\hat{P}_{k},\quad \{\hat{Q}_{i},\hat{J}_{jk}\}=\delta_{ik}\hat{Q}_{j}-\delta_{ij}\hat{Q}_{k}, \\
     \{\hat{J}_{ij},\hat{J}_{kl}\} &=\delta_{il}\hat{J}_{jk}-\delta_{ik}\hat{J}_{jl}+\delta_{jk}\hat{J}_{il}-\delta_{jl}\hat{J}_{ik} .
     \end{split}
\end{equation}
Where the Hamiltonian is central in this algebra. In the above algebra, we do not have any boost symmetry, if we ignore $\hat{Q}$ and $\hat{Q}_{i}$, our algebra reduces to the Aristotelian algebra \cite{Armas:2023ouk}. The background geometry in the absence of boost symmetry is referred to as absolute or Aristotelian space-time. The fracton algebra \eqref{Fracton-alg} can be compared with the Carroll algebra \eqref{carroll-algebra}, where the roles of the boost generator \(\hat{B_i}\) and the dipole generator \(\hat{Q}_{i}\) have been interchanged, as well as the roles of \(\hat{E}\) and \(\hat{Q}\).
\subsection{Fracton-Maxwell algebra}\label{Fracton-Maxwell-algebra}
In this section, we aim to extend fracton algebra to include gauge fields, thereby facilitating a comparison with the Carroll-Maxwell algebra. In the following, we present the \textit{Fracton-Maxwll} algebra.
Considering the following symmetry generators in scalar charge theory, 
\begin{align}\label{def:operatorSFAL}
    \hat{H}=E-d^{i}\partial_{i}A_{0},\quad \hat{P}_{i}=p_{i}-d^{j}A_{ij},\quad\hat{J}_{ij}=\epsilon_{ijk}\epsilon^{lmk}x_{m}(p_{l}-d^{n}A_{ln}),
\end{align}
where the dipole moment is $\vec{d}=q\vec{x}$, the Fracton-Maxwell algebra can be expressed as:
\begin{equation}\label{frac-Max algebra}
\begin{split}
      \{\hat{P}_{i}.\hat{Q}_{j}\} &=-\delta_{ij}\hat{Q}, \quad
      \{\hat{J}_{ij},\hat{P}_{k}\}=\delta_{jk}\hat{P}_{i}-\delta_{ik}\hat{P}_{j} , \\
\{\hat{J}_{ij},\hat{Q}_{k}\} &=\delta_{jk}\hat{Q}_{i}-\delta_{ik}\hat{Q}_{j} , \quad   \{\hat{J}_{ij},\hat{J}_{kl}\} =\delta_{ik}\hat{J}_{jl}-\delta_{il}\hat{J}_{jk}-\delta_{jk}\hat{J}_{il}+\delta_{jl}\hat{J}_{ik}, \\
    \{\hat{P}_{l},\hat{P}_{m}\} &=\hat{Q}^{j}\hat{Z}_{lmj}, \quad \{\hat{H},\hat{P}_{l}\} =\hat{Q}\hat{Z}_{l}+\hat{Q}^{i}\hat{Z}_{il},\\
    \{\hat{J}_{ij},\hat{Z}_{klm}\} &=\delta_{jk}\hat{Z}_{ilm}-\delta_{ik}\hat{Z}_{jlm}+\delta_{jl}\hat{Z}_{kim}-\delta_{il}\hat{Z}_{kjm}+\delta_{jm}\hat{Z}_{kli}-\delta_{im}\hat{Z}_{klj}\\
    \{\hat{J}_{ij},\hat{Z}_{lm}\} &=\delta_{jl}\hat{Z}_{im}-\delta_{il}\hat{Z}_{jm}+\delta_{jm}\hat{Z}_{li}-\delta_{im}Z_{lj}\\
    \{\hat{J}_{ij},\hat{Z}_{m}\} &=\delta_{im}\hat{Z}_{j}-\delta_{jm}\hat{Z}_{i} \, ,
    \end{split}
\end{equation}
where the relations \( Z_{ijk} = \partial_{i} A_{jk} - \partial_{j} A_{ik} \), \( \hat{Z}_{il} = \partial_{t}A_{ij} -\partial_{l}\partial_{i}A_{0} \) and  \( Z_{i} = -\partial_{i} A_{0} \) are defined. The topic of fractons in an electromagnetic field is a highly specialized one that lies at the intersection of quantum field theory, condensed matter physics, and electromagnetism. 

\subsection{Vector charge algebra}\label{Vector-charge-algebra}
As suggested by the vector charge theory of fractons (see Sec.~\ref{section:vector-charge}), we can extend the gauge theory of fractons by replacing the scalar charge density with a vector. Consequently, we identify the operators in vector charge theory as follows:
\begin{align}
    \hat{H}=E,\quad\hat{P}_{i}=p_{i},\quad \hat{Q}_{i}=q_{i},\quad \hat{J}=\vec x\times \vec p,\quad \hat{D}=\vec x\times \vec q.
\end{align}
The operator \(\hat{Q}_{i} \) represents a vector charge operator and the operator \(\hat{D}_{ij} \) denotes a generalized higher-rank dipole moment in vector charge theory.
Hence, the algebra of vector charges can be represented as follows:
\begin{equation} \label{vector charge algebra}
  \begin{split} 
    \{\hat{Q}_{i},\hat{D}_{jk}\} &= \{\hat{Q}_{i},\hat{P}_{j}\}=0,\quad \{\hat{P_i},\hat{D}_{jk}\}=\delta_{ij}\hat{Q}_{k}-\delta_{ik}\hat{Q}_{j} , \\
     \{\hat{P_i},\hat{J}_{jk}\} &=\delta_{ik}\hat{P_j}-\delta_{ij}\hat{P_k},\quad \{\hat{J}_{ij}\,\hat{D}_{kl}\}=\delta_{il}\hat{D}_{jk}-\delta_{ik}\hat{D}_{jl}+\delta_{jk}\hat{D}_{il}-\delta_{jl}\hat{D}_{ik}, \\
     \{\hat{J}_{ij},\hat{J}_{kl}\} &=\delta_{il}\hat{J}_{jk}-\delta_{ik}\hat{J}_{jl}+\delta_{jk}\hat{J}_{il}-\delta_{jl}\hat{J}_{ik} .
     \end{split}
\end{equation}
\subsubsection{Vector charge-Maxwell algebra} 
When a vector charge theory interacts with a symmetric tensor gauge field, $A_{ij}$, certain operators in the vector charge theory acquire additional terms. This can be demonstrated as follows:
\begin{align}
    \hat{H}=E-q^{i}\phi_{i},\quad \hat{P}_{i}=p_{i}-q^{j}A_{ij},\quad\hat{J}_{ij}=\epsilon_{ijk}\epsilon^{lmk}x_{m}(p_{l}-q^{n}A_{ln}).
\end{align}
Here, \(\vec{q}\) represents a vector charge. The above relation can be compared with the definitions of scalar theory \eqref{def:operatorSFAL}. These definitions are supported by the dynamical considerations performed in Secs. (\ref{sec:scalartheory}) and (\ref{section:vector-charge}). Then, the vector charge-Maxwell algebra can be expressed as follows:
\begin{equation} 
  \begin{split} 
     \{\hat{Q}_{i},\hat{D}_{jk}\} &= \{\hat{Q}_{i},\hat{P}_{j}\}=0,\quad \{\hat{P_i},\hat{D}_{jk}\}=\delta_{ij}\hat{Q}_{k}-\delta_{ik}\hat{Q}_{j} , \\
     \{\hat{P_i},\hat{J}_{jk}\} &=\delta_{ik}\hat{P_j}-\delta_{ij}\hat{P_k},\quad \{\hat{J}_{ij}\,\hat{D}_{kl}\}=\delta_{il}\hat{D}_{jk}-\delta_{ik}\hat{D}_{jl}+\delta_{jk}\hat{D}_{il}-\delta_{jl}\hat{D}_{ik}, \\
     \{\hat{J}_{ij},\hat{J}_{kl}\} &=\delta_{il}\hat{J}_{jk}-\delta_{ik}\hat{J}_{jl}+\delta_{jk}\hat{J}_{il}-\delta_{jl}\hat{J}_{ik}, \\
    \{\hat{P_l},\hat{P_m}\} &= \hat{Q}^{j}\hat{Z}_{lmj} , \quad  \{\hat{H},\hat{P_j}\} = \hat{Q}^{i}\hat{Z}_{ij}
    \\
    \{\hat{J}_{ij},\hat{Z}_{klm}\} &=\delta_{jk}\hat{Z}_{ilm}-\delta_{ik}\hat{Z}_{jlm}+\delta_{jl}\hat{Z}_{kim}-\delta_{il}\hat{Z}_{kjm}+\delta_{jm}\hat{Z}_{kli}-\delta_{im}\hat{Z}_{klj}\\
     \{\hat{J}_{ij},\hat{Z}_{lm}\} &=\delta_{jl}\hat{Z}_{im}-\delta_{il}Z_{jm}+\delta_{jm}\hat{Z}_{li}-\delta_{im}\hat{Z}_{lj}.
     \end{split}
\end{equation}
This algebra is closed and has a similar structure to those of the fracton-Maxwell algebra with definitions \( \hat{Z}_{ijk} = \partial_{i} A_{jk} - \partial_{j} A_{ik} \), \( \hat{Z}_{il} = \partial_{t}A_{ij} -\partial_{l}\phi_i \).

\subsection{Algebraic duality}
In this subsection, we will explore the duality between non-Lorentzian and fracton algebras and their extensions.
As explored and demonstrated in Sec.~\ref{FNLduality}, fractons and Non-Lorentzian particles exhibit a duality. We expect to see this duality reflected at the algebraic level. In the following, we explore deeper into their extensions and dualities.

\paragraph{Fracton/Carroll algebra duality :} The fracton algebra \eqref{Fracton-alg} is dual to the Carroll algebra \eqref{carroll-algebra} if the dipole operator $\hat{Q}_{i}$ in the fracton algebra \eqref{Fracton-alg} is replaced with the Carroll boost operator $\hat{B}_{i}$ in the Carroll algebra \eqref{carroll-algebra}. This demonstrates that in both algebras \eqref{Fracton-alg} and \eqref{carroll-algebra}, the Hamiltonian operator $\hat{H}$ serves as a central charge.

\paragraph{Fracton-Maxwell/Carroll-Maxwell algebra duality :} 
The fracton-Maxwell algebra \eqref{frac-Max algebra} describes the symmetries of fracton dipoles in the presence of the symmetric tensor gauge fields $A_{ij}$ and their conjugate fields $E_{ij}$ and ${B}_i^j$. On the other hand, the Carroll-Maxwell algebra \eqref{Carroll-Maxwell} describes the symmetries of Carroll charge particles in the gauge field $\vec{A}$ and the conjugate fields $\vec{E}$ and $\vec{B}$. As we observed at the action and dynamics levels, these particles exhibit duality. However, at the algebraic level, the structures in \eqref{frac-Max algebra} and \eqref{Carroll-Maxwell} show fundamental differences between these algebras, especially for the commutators \(\{\hat{P}_l,\hat{P}_m\}\) and \(\{\hat{H},\hat{P}_m\}\), although they still share some similarities.
\section{Gauging Fracton Algebra and COM Gravity} \label{Fracton Gravity and Non-Lorentzian Gravity0}
As demonstrated, fractons naturally couple to spatial two-index symmetric tensor gauge fields $A_{ij}$. This suggests that dipole conservation is related to spin-2 fields and possibly gravity \cite{Afxonidis:2023pdq,Pretko:2017fbf}. Several authors have investigated the internal dynamics of fracton systems and explored their potential connection to emergent gravitational phenomena. The initial clue arises from the fact that gravity is also described by a symmetric tensor represented by the metric $g_{\mu\nu}$ \cite{Pretko:2017fbf}.
They show that the tensor gauge field $A_{ij}$ is analogous to the spatial components metric tensor, implying a relationship between the conservation of dipoles and the spin-2 fields, and possibly the gravitational field \cite{Afxonidis:2023pdq}. Fractons act as emergent sources for this tensor gauge field, so it can be suggested that fractons generate a non-zero curvature in a manner analogous to how matter acts as a source for the curvature of the metric in gravity. In \cite{Pretko:2017fbf}, it was proposed that the fracton (im)mobility could be understood as a manifestation of Mach's principle. Another interesting connection between fractons and gravity was illustrated in \cite{Afxonidis:2023pdq,Bertolini:2023juh}.
A clear distinction between the tensor gauge theories discussed so far and gravity is that the former are not Lorentz invariant; often, they are not even formulated in a Lorentz covariant manner \cite{Afxonidis:2023pdq}. A deeper connection between fracton order and geometry is also suggested by further studies of lattice models \cite{Slagle:2017mzz,Gu:2006vw,Gu:2009jh}.\\
In \cite{Bertolini:2023juh,Blasi:2022mbl}, they consider the theory of a symmetric tensor field in 4D, invariant under a subclass of infinitesimal diffeomorphism transformations. The most
general action of this theory contains, but is not limited to, linearized gravity. The resulting gauge fixed theory contains both
fractons and linearized gravity as a limit.
In \cite{Afxonidis:2023pdq}, the authors introduce a spacetime Lorentz covariant version of dipole symmetry. By considering gauge fields in the presence of a suitable background field, they develop a theory that includes both a massive gauge-invariant antisymmetric field and a massless symmetric field, reminiscent of linearized gravity.
The work of Bertolini et al. \cite{Bertolini:2022ijb} proposes a generalized covariant Maxwell-like action for the fractons. Starting with a partially symmetric rank-3 tensor field strength $F_{\mu\nu\rho}(x)$ that satisfies a Bianchi identity, they construct an action which remains invariant under the covariant fracton transformation ($\delta_{fracton}A_{\mu\nu}=\partial_{\mu}\partial_{\nu}\Lambda$), comprises two distinct terms: one associated with linearized gravity (LG) and the other specifically related to fractons.
The fracton electrodynamics can also be obtained from a relativistic higher-dimensional theory upon dimensional reduction \cite{Pena-Benitez:2023aat}. It has also been shown that the background geometry in fracton gauge field theories is subject to constraints arising from additional conservation laws. In particular, \cite{Bidussi:2021nmp} derives such a constraint by enforcing the invariance of a fracton gauge theory coupled to curved space. Moreover, in \cite{Slagle:2018kqf}, a constraint is formulated for the background space geometry by considering dipole conservation over closed loops in curved space. One of the key motivations behind our analysis in this section is to investigate the conditions under which similar constraints emerge when using gauging algebra techniques for spacetime symmetries, marking a departure from previous approaches that primarily reported constraints on spatial backgrounds.

In this section, first, we consider fracton algebra and gauge it to determine the geometry to which the associated fields couple. We subsequently examine the spacetime geometry resulting from this gauging procedure, considering two distinct coupling scenarios between the background structure and the gauge fields. Then, the next subsection introduces a new algebraic structure for spacetime, which we term the Center of Mass (COM) algebra. Motivated by fracton algebra, we gauge this COM algebra to investigate the resulting spacetime structure and the associated gravitational theory.

\subsection{Gauging the Fracton Algebra} \label{Fracton Gravity and Non-Lorentzian Gravity}

This subsection focuses on the fracton symmetry group and its underlying geometry using algebra gauging. Following recent work demonstrating that gauging non-Lorentzian algebras (Galilean and Carroll) leads to non-Lorentzian geometries \cite{Lian:2021udx,Hartong:2015xda,Hartong:2015zia}, we are motivated to gauge the full fracton algebra (Eq.~\eqref{dipole symmetry}) to investigate the geometric background to which fractons couple and to analyze the resulting structure.
\subsubsection{The (co)frame bundle}
We start by defining a geometric structure composed  of a 1-form  $\tau_{\mu}$  (clock-form) and a symmetric covariant tensor $h_{\mu\nu}$  (spatial metric), where neither field has been assigned any local tangent space transformations. The signature of $h_{\mu\nu}$ is $(0,1...,1)$. For a better and clearer understanding of the spatial metric $h_{\mu\nu}$, we can decompose it into vielbeins in the following manner 
\begin{align}
    h_{\mu \nu}=\delta_{ab} e_{\mu}^a e^{b}_{\nu},
\end{align}
where $a = 1, . . . , d$ and $\mu$ takes $d + 1$ values. We can therefore introduce a dual base (or frame) ($v^{\mu}$, $e^{\mu}_{a}$) for ($\tau_{\mu}$, $e^{a}_{\mu}$), so that the following orthonormality relations are satisfied by these 
\begin{align} \label{vielbein}
    v^{\mu}\tau_{\mu}=-1,\quad v^{\mu}e^{a}_{\mu}=e^{\mu}_{a}\tau_{\mu}=0,\quad e^{a}_{\mu}e^{\mu}_{b}=\delta^{a}_{b}.
\end{align}
The spatial inverse metric $h^{\mu\nu}$, can also be defined as follows:
\begin{align}
     h^{\mu \nu}=\delta^{ab}e^{\mu}_{a}e^{\nu}_{b}.
\end{align}
Although this geometry shares a similar structure with Newton-Cartan and Carrollian geometries \cite{Festuccia:2016awg,Hartong:2015zia,Andringa:2010it,Christensen:2013lma,deBoer:2023fnj,Ciambelli:2019lap,Hartong:2015xda,Marsot:2022imf}, it will eventually lead to distinct dynamics. Fracton algebra represents an algebra devoid of boost operators. In geometric terms, we can describe fractons as residing within the framework of spacetime geometry with absolute time. 
\subsubsection{Algebra valued connection and curvatures}
The fracton algebra \eqref{Fracton-alg} contains five generators: spatial translations $\hat{P}_{i}$, time translations $\hat{H}$, rotation $\hat{J}_{ij}$, and also charge and dipole generators $\hat{Q}$ and $\hat{Q}_{i}$.
To construct the vielbein formalism and ultimately the spacetime geometry of fractons, one can directly define the vielbein bundle and its connections. We can see that these vilbines are constructed by gauging the fracton algebra, as the gauging procedure of Carroll and Galilean algebras gives us a compatible vilbine system \cite{Hartong:2015zia, Hartong:2015xda, Lian:2021udx}.

In order to gauge the fracton  algebra we define a connection $A_{\mu}$ as
\begin{align}\label{connection}
 A_{\mu}=\hat{H}\tau_{\mu}+\hat{P}_{a}e^{a}_{\mu}+\frac{1}{2}\hat{J}_{ab}\omega^{ab}_{\mu}+\hat{Q}_{a}n^{a}_{\mu}+\hat{Q}n_{\mu},
\end{align}
the value of $\mu$ assumes $d+1$  distinct values, reflecting the fact that the system contains one time generator and $d$ space translation generators. In general, we work with a $(d+1)$-dimensional space-time. This connection transforms as \cite{Armas:2023ouk}
\begin{align} \label{gauge transformation}
    \delta A_{\mu}=\partial_{\mu}\Pi+[A_{\mu},\Pi],
\end{align}
where we defined $\Pi$ as
\begin{align}
    \Pi=\xi^{\mu}A_{\mu}+\Sigma,
\end{align}
in this context, $\xi^\mu$ can be considered the spacetime parameter, and $\Sigma$ is defined as
\begin{align}\label{rel:internalfields}
    \Sigma=\frac{1}{2}\hat{J}_{ab}\lambda^{ab}+\hat{Q}_{a}\lambda^a+\hat{Q}\lambda,
\end{align}
in which $\lambda^{ab}$, $\lambda^{a}$ and $\lambda$ are the local rotation, local dipole shift and the standard $U(1)$ gauge transformation, respectively \cite{Armas:2023ouk}. Eq.(\ref{gauge transformation}) represents a gauge transformation, but it is not a valid internal transformation at this stage. To obtain a proper internal transformation rule, more precisely, with the help of the curvature $2$-form \cite{Hartong:2015zia, Hartong:2015xda, Lian:2021udx,Armas:2023ouk}
\begin{align}
    F_{\mu\nu}=\partial_{\mu}A_{\nu}-\partial_{\nu}A_{\mu}+[A_{\mu},A_{\nu}],
\end{align}
one can define the proper internal transformation $\Bar{\delta}$ based on $\delta$ as
\begin{align}\label{proper transformation}
    \Bar{\delta}A_{\mu}=\delta A_{\mu}-\xi^{\nu}F_{\mu \nu}=\mathcal{L}_{\xi}A_{\mu}+\partial_{\mu}\Sigma+[A_{\mu},\Sigma].
\end{align}
We can easily derive the proper transformations for our vielbeins in (\ref{connection}) as the following
\begin{align}
    \Bar{\delta}\tau_{\mu} &=\mathcal L_{\xi}\tau_{\mu},\\
\Bar{\delta}e^{a}_{\mu} &=\mathcal L_{\xi}e^{a}_{\mu}+\lambda^{a}_{b}e^{b}_{\mu},\\
\Bar{\delta}n^{a}_{\mu} &=\mathcal L_{\xi}n^{a}_{\mu}+\partial_{\mu}\lambda^{a}+\lambda^{a}_{b}n^{b}_{\mu}+\lambda^{b}\omega^{a}_{\mu \ b}, \\
    \Bar{\delta}n_{\mu} &=\mathcal L_{\xi}n_{\mu}+\partial_{\mu}\lambda+\lambda_{a}e^{a}_{\mu}, \\
    \Bar{\delta}\omega^{ab}_{\mu} &=\mathcal L_{\xi}\omega^{ab}_{\mu}+\partial_{\mu}\lambda^{ab}+\lambda^{a}_{c}\omega^{cb}_{\mu}-\lambda^{b}_{c}\omega^{ca}_{\mu}.
\end{align}
The physical interpretation of the curvature 2-form involved in our local gauge transformation is essential to study. The curvature 2-form projection can be written as
\begin{align}
    F_{\mu\nu}=\hat{H}R_{\mu\nu}(H)+\hat{P}_{a}R_{\mu\nu}{}^{a}(P)+\frac{1}{2}\hat{J}_{ab}R_{\mu\nu}{}^{ab}(J)+\hat{Q}_{a}R_{\mu\nu}{}^{a}(Q_{a})+\hat{Q}R_{\mu\nu}(Q),
\end{align}
then the expressions for the components of the field strength $F_{\mu\nu}$ for the fracton symmetry are as follows:
\begin{align}
    R_{\mu\nu}(H)&=\partial_{[\mu}\tau_{\nu]},\\
    R_{\mu\nu}{}^{a}(P)&=\partial_{[\mu}e^{a}_{\nu]}-\omega^{ab}_{[\mu}e_{\nu]b},\\
    R_{\mu\nu}{}^{a}(Q_{a})&=\partial_{[\mu}n^{a}_{\nu]}-\omega^{ab}_{[\mu}n_{\nu]b},\\
    R_{\mu\nu}(Q)&=\partial_{[\mu}n_{\nu]}-e^{a}_{[\mu}n_{\nu]a},\\
    R_{\mu\nu}{}^{ab}(J)&=\partial_{[\mu}\omega^{ab}_{\nu]}-\omega^{ca}_{[\mu}\omega^{b}_{\nu]c}. 
\end{align}
The curvature 2-form is a fundamental entity that plays a significant role in analyzing the constraints of the base manifold. In Aristotelian geometry, there exist only 2-form curvatures $R_{\mu\nu}(H)$, $R_{\mu\nu}{}^{a}(P)$, and $R_{\mu\nu}{}^{ab}(J)$. 
In fracton algebra, we encounter a greater variety of symmetries compared to Aristotelian symmetry, including dipole and $U(1)$ symmetries. Consequently, fracton geometry exhibits more 2-form curvatures than Aristotelian geometry. However, we consider dipole and $U(1)$ symmetries not as part of spacetime symmetries but rather as internal symmetries, as explained in \cite{Armas:2023ouk}. The curvature 2-forms $R_{\mu\nu}(H)$ and $R_{\mu\nu}{}^{a}(P)$ describe both temporal and spatial torsion, while $R_{\mu\nu}{}^{ab}(J)$ specifically characterizes spatial curvature.

One possible way to define the covariant derivative, denoted by $\mathcal{D}_{\mu}$, in our framework is as follows:
\begin{equation}
\begin{split} \label{def:covariants}
    \mathcal{D}_{\mu}\tau_{\nu} &=\partial_{\mu}\tau_{\nu}-\Gamma^{\rho}_{\mu\nu}\tau_{\rho},\\
    \mathcal{D}_{\mu}e^{a}_{\nu} &=\partial_{\mu}e^{a}_{\nu}-\Gamma^{\rho}_{\mu\nu}e^{a}_{\rho}-\omega^{a}_{\mu b}e^{b}_{\nu}.
\end{split}
\end{equation}
This formulation allows us to characterize the curvature properties in $\mathcal{F}_{\mu \nu}$ using an affine connection $\Gamma^{\rho}_{\mu\nu}$, which must remain invariant under fracton symmetry transformations. Notably, the definition of the covariant derivative excludes any direct coupling between the geometric background and the fracton gauge fields. We start by requiring covariance conditions \cite{Hartong:2015zia, Hartong:2015xda, Lian:2021udx,Armas:2023ouk}:
\begin{align}
   & \mathcal{D}_{\mu}\tau_{\nu}=0,\\
   & \mathcal{D}_{\mu}e^{a}_{\nu}=0.
\end{align}
Thus, we find $\Gamma^{\nu}_{\mu\rho}$ in terms of the vielbeins and connections,
\begin{align}\label{Christoffel symbols 0}
    \Gamma^{\rho}_{\mu\nu}e^{a}_{\rho}e^{\lambda}_{a}&=e^{\lambda}_{a}(\partial_{\mu}e^{a}_{\nu}-\omega^{a}_{\mu b}e^{b}_{\nu}),\\ \Gamma^{\rho}_{\mu\nu}\tau_{\rho}v^{\lambda}&=-v^{\lambda}\partial_{\mu}\tau_{\nu},
\end{align}
where $v^{\lambda}$ and $e^{\lambda}_{a}$ are the temporal and spatial inverse vielbines. By following relation (\ref{vielbein}), we can naturally find the fundamental orthonormality condition
\begin{align}\label{kernel}
    v^{\mu}h_{\mu\nu}=0,
\end{align}
and we can find that the protective inverse of $\tau_{\mu}$ and $h_{\mu\nu}$ naturally satisfy the completeness relations
\begin{align}
    v^{\mu}\tau_{\nu}=-\delta^{\mu}_{\nu}+h^{\mu\lambda}h_{\lambda\nu}.
\end{align}
The inverse vielbein transforms under the $\Bar{\delta}$ transformations as
\begin{align}
    \Bar{\delta}v^{\mu} &=\mathcal{L}_{\xi}v^{\mu}, \\
    \Bar{\delta}e^{\mu}_{a} &=\mathcal{L}_{\xi}e^{\mu}_{a}+\lambda_{a}^{b}e^{\mu}_{b},
\end{align}
then we find
\begin{align}
   \mathcal{D}_{\mu}v^{\nu} &=\partial_{\mu}v^{\nu}+\Gamma^{\nu}_{\mu\rho}v^{\rho}=0, \\
    \mathcal{D}_{\mu}e^{\nu}_{a} &=\partial_{\mu}e^{\nu}_{a}+\Gamma^{\nu}_{\mu\rho}e^{\rho}_{a}-\omega^{\mu b}_{a}e^{\nu}_{b}=0.
\end{align}
By subtracting the equations in (\ref{Christoffel symbols 0})  and employing the vielbein completeness relation, we can derive the following equation
\begin{align}\label{ch-symbol}
    \Gamma^{\rho}_{\mu\nu}=-v^{\rho}\partial_{\mu}\tau_{\nu}+e^{\rho}_{a}(\partial_{\mu}e^{a}_{\nu}-\omega^{a}_{\mu b}e^{b}_{\nu}) \, .
\end{align}
Now, we are ready to find the torsion tensor $\Gamma^{\rho}_{[\mu\nu]}$ and the Riemann curvature tensor $R^{\rho}_{\mu\nu\sigma}$ as follows: 
\begin{align}
    2\Gamma^{\rho}_{[\mu\nu]}&=-2v^{\rho}\partial_{[\mu}\tau_{\nu]}+2e^{\rho}_{a}(\partial_{[\mu}e^{a}_{\nu]}-\omega^{a}_{[\mu |b|}e^{b}_{\nu]})\\&=-v^{\rho} R_{\mu\nu}(H)+e^{\rho}_{a} R_{\mu\nu}{}^{a}(P).
\end{align}
As mentioned in the previous subsection, $R_{\mu\nu}(H)$ and $R_{\mu\nu}{}^{a}(P)$ represent the temporal and spatial components of the spacetime torsion. The Riemann curvature tensor $R^{\rho}_{\mu\nu\sigma}$ defined as 
\begin{align}
    R^{ \rho}_{\mu\nu\sigma}=-\partial_{\mu} \Gamma^{\rho}_{\nu\sigma}+\partial_{\nu} \Gamma^{\rho}_{\mu\sigma}-\Gamma^{\rho}_{\mu\lambda}\Gamma^{\lambda}_{\nu\sigma}+\Gamma^{\rho}_{\nu\lambda}\Gamma^{\lambda}_{\mu\sigma},
\end{align}
thus, the Riemann curvature tensor satisfies 
\begin{align}\label{eq:riemann-nocoupling}
    R^{\rho}_{\mu\nu\sigma}=-e^{\rho}_{b}e_{\sigma \ a}(\partial_{[\mu}\omega^{ab}_{\nu]}-\omega^{ca}_{[\mu}\omega^{b}_{\nu]c})=-e^{\rho}_{b}e_{\sigma \ a}R_{\mu\nu}{}^{ab}(J).
\end{align}

\subsubsection{Aristotelian  Affine Connection }
In this subsection we aim to find the general solution for the metric connection. Starting with the vielbein postulate and metric compatibility, in fracton case by utilizing the fact that $\omega^{ab}_{\mu}$ is antisymmetric, we can deduce that compatibility with the fracton structure implies that
\begin{align} \label{compatibility condition}
    \nabla_{\rho}h_{\mu\nu}=\nabla_{\rho}h^{\mu\nu}
    =0.
\end{align}
The most general form of $\Gamma^{\nu}_{\mu\rho}$ obeys the condition of metric compatibility, which states ``$\nabla_{\rho}h_{\mu\nu}=\nabla_{\mu}h_{\rho\nu}=\nabla_{\nu}h_{\mu\rho}=0$''. From vielbein postulate of $e^{a}_{\mu}$ we have 
\begin{align}\label{condition}
    \nabla_{\mu}h_{ \rho\nu}=\partial_{\mu}h_{\rho\nu}-\Gamma^{\lambda}_{\mu\nu}h_{\rho\lambda}-\Gamma^{\lambda}_{\mu\rho}h_{\nu\lambda}=0.
\end{align}
Through the application of permutations to the indices, followed by the summation of the resulting values, the desired result is obtained as follows:
\begin{align}\label{Christoffel symbols 1}
    2\Gamma^{\lambda}_{(\mu\nu)}h_{\lambda\rho}=(\partial_{\mu}h_{\nu\rho}+\partial_{\nu}h_{\mu\rho}-\partial_{\rho}h_{\mu\nu})-2\Gamma^{\lambda}_{[\mu\rho]}h_{\nu\lambda}-2\Gamma^{\lambda}_{[\nu\rho]}h_{\mu\lambda},
\end{align}
this equation can be contracted with $v^{\rho}$, which gives
\begin{align}\label{the extrinsic curvature}
    K_{\mu\nu}=-v^{\rho}h_{\nu\lambda}\Gamma^{\lambda}_{[\rho\mu]}-v^{\rho}h_{\mu\lambda}\Gamma^{\lambda}_{[\rho\nu]},
\end{align}
while the extrinsic curvature, $K_{\mu\nu}$, is defined as:
\begin{align}
    K_{\mu\nu}=-\frac{1}{2}\mathcal{L}_{v}h_{\mu\nu},
\end{align}
which is a purely spatial quantity, $v^{\mu}K_{\mu\nu}=0$. In accordance with the definition of the extrinsic curvature provided in equation (\ref{the extrinsic curvature}), we propose the following ansatz:
\begin{align} \label{Christoffel symbols 2}
    \Gamma^{\lambda}_{[\rho\mu]}=\tau_{[\rho}K_{\mu]\sigma}h^{\lambda\sigma}+\Theta^{\lambda}_{[\rho\mu]},
\end{align}
where $\Theta^{\lambda}_{[\rho\mu]}$ is an arbitrary tensor. This ansatz is motivated by \cite{Hartong:2015xda}, where Carroll algebra is studied. It differs from those considered in \cite{Bidussi:2021nmp} and \cite{deBoer:2020xlc}.
The conditions in (\ref{condition}) remain valid if we have
\begin{align}
    v^{\mu}\Theta^{\lambda}_{[\rho\mu]}=0, \quad\Theta^{\lambda}_{[\rho\mu]}h_{\nu\lambda}+\Theta^{\lambda}_{[\mu\nu]}h_{\rho\lambda=0},
\end{align}
assuming that $\Theta^{\lambda}_{[\rho\mu]}$ is entirely spatial.
Now, by considering (\ref{Christoffel symbols 2}) and using this relation in (\ref{Christoffel symbols 1}), and then by adding $2\Gamma^{\lambda}_{[\mu\nu]}h_{\lambda\rho}$ to both sides of (\ref{Christoffel symbols 1}), we obtain the following relation for the connection:
\begin{equation}\label{Christoffel symbols 3}
\begin{split}
    2\Gamma^{\lambda}_{\mu\nu}h_{\lambda\rho} =&(\partial_{\mu}h_{\nu\rho}+\partial_{\nu}h_{\mu\rho}-\partial_{\rho}h_{\mu\nu})+2\tau_{[\rho}K_{\mu]\sigma}h^{\lambda\sigma}h_{\nu\lambda}+2\tau_{[\rho}K_{\nu]\sigma}h^{\lambda\sigma}h_{\mu\lambda} \\
       &+2\tau_{[\nu}K_{\mu]\sigma}h^{\lambda\sigma}h_{\rho\lambda} +\Theta^{\lambda}_{[\rho\mu]}h_{\nu\lambda}+\Theta^{\lambda}_{[\rho\nu]}h_{\mu\lambda}+\Theta^{\lambda}_{[\rho\mu]}h_{\rho\lambda}.
       \end{split}
\end{equation}
The Christoffel symbol above is analogous to the Carrollian Christoffel symbol discussed in \cite{Hartong:2015xda}. However, a fundamental difference exists between them: The Christoffel symbol in (\ref{Christoffel symbols 3}) is invariant under Aristotelian transformations, unlike the Carrollian Christoffel symbol presented in \cite{Hartong:2015xda}.

We can further simplify equation (\ref{Christoffel symbols 3}) by multiplying both sides by \(h^{\rho\sigma}\) and using the fact that \(\nabla_{\mu}\tau_{\nu}=0\). Then, we find
\begin{align}
\begin{split}
    \Gamma^{\sigma}_{\mu\nu}=&-v^{\sigma}\partial_{\mu}\tau_{\nu}-\frac{1}{2}h^{\rho\sigma}(\partial_{\mu}h_{\nu\rho}+\partial_{\nu}h_{\mu\rho}-\partial_{\rho}h_{\mu\nu})-h^{\rho\sigma}\tau_{\nu}K_{\mu\rho}\\&+\frac{1}{2}h^{\rho\sigma}(2\Theta^{\lambda}_{[\rho\mu]}h_{\nu\lambda}+2\Theta^{\lambda}_{[\rho\nu]}h_{\mu\lambda}+2\Theta^{\lambda}_{[\rho\mu]}h_{\rho\lambda}).
\end{split}
\end{align}
We can define the spatial tensor $S_{\rho\mu\nu}$
\begin{align}
    S_{\rho\mu\nu}=2\Theta^{\lambda}_{[\rho\mu]}h_{\nu\lambda}+2\Theta^{\lambda}_{[\rho\nu]}h_{\mu\lambda}+2\Theta^{\lambda}_{[\rho\mu]}h_{\rho\lambda},
\end{align}
where $v^{\mu}S_{\rho\mu\nu}=0$.
Finally, we are able to present the definition of an affine connection in Aristotelian geometry as follows:
\begin{align}\label{Christoffel symbols 4}
    \Gamma^{\sigma}_{\mu\nu}=-v^{\sigma}\partial_{\mu}\tau_{\nu}-\frac{1}{2}h^{\rho\sigma}(\partial_{\mu}h_{\nu\rho}+\partial_{\nu}h_{\mu\rho}-\partial_{\rho}h_{\mu\nu})+h^{\rho\sigma}(-\tau_{\nu}K_{\mu\rho}+\frac{1}{2}S_{\rho\mu\nu}).
\end{align}
This connection incorporates torsion through two distinct terms: an intrinsic torsion, represented by the expressions \( \partial_{[\mu} \tau_{\nu]} \) and \( K_{\mu\nu} \), and a contribution from \( S^{\sigma}_{[\mu\nu]} \), which is defined as follows:
\begin{align}
     \Gamma^{\sigma}_{[\mu\nu]}=-v^{\sigma}\partial_{[\mu}\tau_{\nu]}-h^{\rho\sigma}\tau_{[\nu}K_{\mu]\rho}+\frac{1}{2}S^{\sigma}_{[\mu\nu]} \, .
\end{align}
If we set \( S^{\sigma}_{[\mu\nu]} = 0 \), then the Aristotelian geometry possesses only an intrinsic torsion that can be represented by the expressions $\partial_{[\mu}\tau_{\nu]}$ and $K_{\mu\nu}$ \cite{Figueroa-OFarrill:2020gpr}. The condition (\ref{compatibility condition}) implies that \( S^{\sigma}_{\mu\nu}v^{\mu} = 0 \) and \( S^{\sigma}_{\mu\nu}v^{\mu}\tau_{\sigma} = 0 \), leading to the result \( h^{\sigma\rho}S_{\rho\mu\nu} = S^{\sigma}_{\mu\nu} \).
Thus, the final expression for the connection follows from imposing intrinsic torsion and metric compatibility of the affine connection. As a result, we obtain:
\begin{align}\label{Christoffel symbols 5}
    \Gamma^{\sigma}_{\mu\nu}=-v^{\sigma}\partial_{\mu}\tau_{\nu}+\frac{1}{2}h^{\rho\sigma}(\partial_{\mu}h_{\nu\rho}+\partial_{\nu}h_{\mu\rho}-\partial_{\rho}h_{\mu\nu})-h^{\rho\sigma}\tau_{\nu}K_{\mu\rho}.
\end{align}
The requirement that \( \Gamma^{\sigma}_{\mu\nu} \) transforms as an affine connection under general coordinate transformations while remaining invariant under dipole symmetry (or fracton symmetry) is satisfied by expression \eqref{Christoffel symbols 5}.
By combining equations \eqref{ch-symbol} and \eqref{Christoffel symbols 4}, we can express \( S_{\rho\mu\nu} \) in terms of geometric quantities as follows:
\begin{align}
    \frac{1}{2}h^{\rho\sigma}S_{\rho\mu\nu}=-\frac{1}{2}h^{\rho\sigma}(\partial_{\mu}h_{\nu\rho}+\partial_{\nu}h_{\mu\rho}-\partial_{\rho}h_{\mu\nu})+h^{\rho\sigma}\tau_{\nu}K_{\mu\rho}+e^{\rho}_{a}\partial_{\mu}e^{a}_{\nu}-\omega^{ a}_{\mu b}e^{\rho}_{a}e^{b}_{\nu}.
\end{align}
In the following subsection, we will introduce a new contravariant vector field to determine the tensor \( S_{\rho\mu\nu} \).
\subsubsection{Redefinition of fields and affine connection}
In this section, we will examine the $\bar{\delta}$ transformations of the set of fields $\tau_{\mu}$, $e^{a}_{\mu}$, $n_{\mu}$, $n^{a}_{\mu}$, and $\omega^{ab}_{\mu}$ in order to reduce the number of independent fields in the given theory. In the context of gauging algebras, such as the Poincaré algebra, it becomes feasible to apply the $\bar \delta$ transformations to a reduced set of fields by imposing curvature constraints. For example, setting the torsion to zero enables us to express the spin connection coefficients $\omega^{ab}_{\mu}$, in terms of the tetrad components $e^{a}_{\mu}$, as demonstrated in reference \cite{Hartong:2015xda}. In the context of fracton algebra, no geometric principles allow us to set $R_{\mu\nu}(Q)$ to zero. However, a selection of novel spatial and temporal vielbein combinations provides an opportunity to establish a geometric framework, which reduces the set of independent fields.

We can select new spatial and temporal vielbein combinations within the fracton algebra that will be invariant under fracton symmetry. For this purpose, we can introduce a contravariant vector field $n^{\mu}$ which transform as
\begin{align}
    \bar\delta n^{\mu}=\mathcal{L}_{\xi}n^{\mu}.
\end{align}
This allows us to construct dipole-invariant vielbein combinations such as the following
\begin{align}\label{New vielbein}
    \hat{e}^{\mu}_{a}&=e^{\mu}_{a}-n^{\sigma}e_{ \sigma a}v^{\mu},\\ \hat{\tau}_{\mu}&=\tau_{\mu}-h_{\mu\sigma}n^{\sigma},
\end{align}
that satisfy
\begin{align}
    v^{\mu}\hat{\tau}_{\mu}=-1,\quad v^{\mu}e^{a}_{\mu}=0,\quad \hat{e}^{\mu}_{a}\hat{\tau}_{\mu}=0,\quad e^{a}_{\mu}\hat{e}^{\mu}_{b}=\delta^{a}_{b}.
\end{align}
Now we can define two dipole-invariant metric-like quantities. The first invariant metric can be obtained as
\begin{align}\label{metric 1}
    \bar{h}^{\mu\nu}=h^{\mu\nu}-n^{\mu}v^{\nu}-n^{\nu}v^{\mu},
\end{align}
where the second invariant is obtained when $\hat{e}^{\mu}_{a}$ is applied to a different metric
\begin{align}\label{metric 2}
    \hat{h}^{\mu\nu}=\delta^{ab}\hat{e}^{\mu}_{a}\hat{e}^{\nu}_{b}=\bar{h}^{\mu\nu}+\alpha\bar\Phi v^{\mu}v^{\nu},
\end{align}
in which we define $\bar\Phi$ as follows:
\begin{align}
    \bar\Phi=-n^{\nu}\tau_{\nu}+\frac{1}{2}h_{\nu\sigma}n^{\nu}n^{\sigma}.
\end{align}
These two dipole invariant metric must satisfy the fundamental orthonormality condition $\hat{\tau}_{\mu}\hat{h}^{\mu\nu}=0$.\\
In the previous subsection, we proceeded to find the general solution for the metric connection, starting with the vielbein postulates and the metric compatibility, from which we find $\Gamma^{\rho}_{\mu\nu}$ in terms of $\tau_{\mu}$, ${e}_{\mu}^{a}$, and $h_{\mu\nu}$. Here, introducing the vector field $n^{\mu}$ further motivates us to express $\Gamma^{\rho}_{\mu\nu}$ in terms of $n^{\mu}$ as well. Our primary motivations can be summarized into two key objectives. Firstly, there is no compelling reason to set \( R_{\mu\nu}(Q) = 0 \); therefore, we accept this as torsion. Secondly, this approach provides a straightforward set of solutions for metric connections concerning the vector \( n^{\mu} \). To achieve this, we must determine the tensor $S_{\rho\mu\nu}$ in terms of $\tau_{\mu}$, ${e}_{\mu}^{a}$, and $n^{\mu}$ and ensure that they are invariant under the transformations $\bar\delta$. Since $S_{\rho\mu\nu}$ is a completely spatial tensor, we have
\begin{align}\label{tensor}
    S_{\rho\mu\nu}=2h_{\nu\lambda}n^{\lambda}K_{\mu\rho}-2h_{\rho\lambda}n^{\lambda}K_{\nu\mu},
\end{align}
which is invariant under dipole symmetry.\\
By considering (\ref{tensor}) and spatial metric (\ref{metric 1}) and substituting them into (\ref{Christoffel symbols 4}), we get
\begin{align} \label{affine connection 1}
    \Gamma^{\sigma}_{\mu\nu}=-v^{\sigma}\partial_{\mu}\hat{\tau}_{\nu}+\frac{1}{2}\bar{h}^{\rho\sigma}(\partial_{\mu}h_{\nu\rho}+\partial_{\nu}h_{\mu\rho}-\partial_{\rho}h_{\mu\nu})-\bar{h}^{\rho\sigma}\hat{\tau}_{\nu}K_{\mu\rho}+\bar{h}^{\rho\sigma}\hat{\tau}_{\rho}K_{\mu\nu}.
\end{align}
The affine connection (\ref{affine connection 1}) possesses the property that when we substitute the metric $\bar{h}^{\rho\sigma}$ with the metric $\hat{h}^{\rho\sigma}$, the expression for $\Gamma^{\sigma}_{\mu\nu}$ remains unchanged. Taking $\alpha=2$, we find
\begin{align} \label{affine connection 2}
    \Gamma^{\sigma}_{\mu\nu}=-v^{\sigma}\partial_{\mu}\hat{\tau}_{\nu}+\frac{1}{2}\bar{h}^{\rho\sigma}(\partial_{\mu}h_{\nu\rho}+\partial_{\nu}h_{\mu\rho}-\partial_{\rho}h_{\mu\nu})-\hat{h}^{\rho\sigma}\hat{\tau}_{\nu}K_{\mu\rho}.
\end{align}
The connection in (\ref{affine connection 1}) and (\ref{affine connection 2}) satisfies the metric compatibility conditions $ \nabla_{\rho}h_{\mu\nu}=\nabla_{\mu}v^{\nu}=0$. However, there are conditions that these connections must satisfy, which are as follows
\begin{align}
    \nabla_{\rho}\hat{\tau}_{\nu}=\nabla_{\rho}\hat{h}^{\mu\sigma}=0.
\end{align}
Before introducing the vector $n^{\mu}$ in the previous subsection, it was evident that the torsion tensors could not be determined uniquely. However, with the inclusion of the vector $n^{\mu}$, we can now fix the torsion tensor. Thus, we arrive at the following relation
\begin{align} \label{torsion}
    \Gamma^{\rho}_{[\mu\nu]}={\hat e}^{\rho}_{a}R_{\mu\nu}{}^{a}(P).
\end{align}
Now we can fix the temporal torsion by utilizing the equation (\ref{torsion}) as follows
\begin{align}
    R_{\mu\nu}(H)=-\hat{\tau}_{\rho}\Gamma^{\rho}_{[\mu\nu]}=n_{a}R_{\mu\nu}{}^{a}(P),
\end{align}
where $n_{a}=e_{\rho\ a}n^{\rho}$. It seems that we can introduce another set of definitions by introducing an additional vielbein field \( m_{\rho} \) for this case, as we know that Aristotelian geometry admits many invariants. Thus, we can present the new definition as follows:
\begin{align}\label{New veilbeine 2}
    \hat{e}_{\mu}^{a}&=e_{\mu}^{a}-e^{\rho a}m_{\rho}\tau_{\mu},\\ \hat{v}^{\mu}&=v^{\mu}-h^{\mu \rho}m_{\rho},
\end{align}
that this vielbeine field $m_{\mu}$ transforms as 
\begin{align}
    \bar\delta m_{\mu}=\mathcal{L}_{\xi}m_{\mu}.
\end{align}
Our work differs from previous studies \cite{Armas:2023ouk}, \cite{Pena-Benitez:2023aat}, and \cite{Afxonidis:2023pdq}. In \cite{Armas:2023ouk}, the authors utilize fracton algebra to extract both the background geometry and the low-energy degrees of freedom, which are important for hydrodynamics and for describing field theories with a conserved dipole moment. We extended their procedure to construct the Riemann curvature and demonstrated that we can reduce the number of free fields in the theory by constructing a new vielbein. In \cite{Pena-Benitez:2023aat}, the authors investigate the relationship between symmetric gauge fields and gravity, building upon their understanding of MDMA (Monopole Dipole Momentum Algebra) as a contraction of the Poincaré algebra. This analysis offers a geometric interpretation of the fracton charge as the momentum of the matter field in a transverse (internal) spacetime dimension, while the dipole charge is interpreted as the angular momentum in that direction. In \cite{Afxonidis:2023pdq}, the authors explore the Lorentz-covariant generalization of gauge theories, in which both an Abelian charge and its associated spacetime are conserved. Generally, the resulting theory encompasses a massive gauge-invariant antisymmetric field and a massless symmetric field. The latter transforms under gauge transformations as a generalization of gravity. In this subsection, we employed the gauging algebra procedure to analyze the fracton algebra and its geometric implications. We found no constraint on the background geometry, due to the absence of coupling between the geometric structure and fracton fields, as indicated in Eq.~\eqref{def:covariants}. This can be further examined through the Ricci scalar $R = e_{a}^{\mu}e_{b}^{\nu}R_{\mu\nu}{}^{ab}(J)$ derived from the Riemann tensor in Eq.~\eqref{eq:riemann-nocoupling}, which remains explicitly invariant under dipole symmetry. This approach allows us to derive geometric entities and establish a framework for fractonic theories. 
\subsubsection{Coupling geometry with gauge field}\label{sec:couplingtogeometry}
Instead of the definitions provided in Eq.~\eqref{def:covariants}, an alternative formulation of the covariant derivative under dipole symmetry can be introduced, wherein the coupling to fracton gauge fields is incorporated directly into its structure as follows:
\begin{align}\label{def:covariants2}
    \tilde{\mathcal{D}}_{\mu}\tau_{\nu} &=\partial_{\mu}\tau_{\nu}-\Gamma^{\rho}_{\mu\nu}\tau_{\rho}-n_{\mu}^{a}e_{\nu}{}_{a}=0,\\
    \tilde{\mathcal{D}}_{\mu}e^{a}_{\nu} &=\partial_{\mu}e^{a}_{\nu}-\Gamma^{\rho}_{\mu\nu}e^{a}_{\rho}-\omega^{a}_{\mu b}e^{b}_{\nu}=0,
\end{align}
A tilde over quantities indicates that they are evaluated in the modified framework, where a coupling between the background geometry and the fracton gauge field is present. We, then, have the following form for the connection:
\begin{align}
    \tilde{\Gamma}_{\mu\nu}^{\sigma}=-v^{\sigma}(\partial_{\mu}\tau_{\nu}-n_{\mu}^{a}\ e_{\nu\ a})+e^{\sigma}_{a}(\partial_{\mu}e_{\nu}^{a}-\omega^{\ a}_{\mu\ b}\ e_{\nu}^{b}) \, ,
\end{align}
from which the Riemann curvature tensor can be obtained as 
\begin{align}
    \tilde{R}_{\mu\nu\sigma}{}^{\rho}=-v^{\rho} e_{\sigma\ a}R_{\mu\nu}{^{a}}(Q_b)-e^{\rho}_{b}e_{\sigma \ a}R_{\mu\nu}{}^{ab}(J).
\end{align}
Thus, the curvature scalar takes the form
\begin{equation}
    \tilde{R} = 2v^{\mu}e_{a}^{\nu}R_{\mu\nu}{}^{a}(Q_b)+e_{a}^{\mu}e_{b}^{\nu}R_{\mu\nu}{}^{ab}(J) \, ,
\end{equation}
where the first term is generally not invariant under dipole symmetry transformations. The transformation properties of the curvature components \( R_{\mu\nu}{}^{a}(Q_b) \) and \( R_{\mu\nu}{}^{ab}(J) \) under dipole symmetry are as follows:
\begin{align}
   \Bar \delta R_{\mu\nu}{}^{a}(Q_b)&=\lambda^{a}{}_{b}R_{\mu\nu}{}^{b}(Q_b) -\lambda_{b}R_{\mu\nu}{}^{ab}(J)\\
    \Bar \delta R_{\mu\nu}{}^{ab}(J)&=\lambda^{b}{}_{c}R_{\mu\nu}{}^{ca}(J)-\lambda^{a}{}_{c}R_{\mu\nu}{}^{cb}(J) \, ,
\end{align}
where we only looked at the dipole part of $\Bar \delta$ transformations. The variation of the Ricci scalar $ \tilde{R}$ under dipole symmetry then yields the following expression:
\begin{align}
     \bar \delta \tilde{R}&=\left(2v^{\mu} \lambda_{a}{}^{b}e^{\nu}_{b}\ R_{\mu\nu}{}^{a}(Q_b)+2v^{\mu}e_{a}^{\nu}(\lambda^{a}{}_{b}R_{\mu\nu}{}^{b}(Q_b) -\lambda_{b}R_{\mu\nu}{}^{ab}(J))\right) \nonumber \\
     &\quad+ \left((\lambda_{a}{}^{c}e^{\mu}_{c}e^{\nu}_{b}+e^{\mu}_{a}\lambda_{b}{}^{c}e^{\nu}_{c})R_{\mu\nu}{}^{ab}(J)+e^{\mu}_{a}e^{\nu}_{b}(\lambda^{b}{}_{c}R_{\mu\nu}{}^{ca}(J)-\lambda^{a}{}_{c}R_{\mu\nu}{}^{cb}(J)) \right) \nonumber \\
     &= -2v^{\mu}e_{a}^{\nu}\lambda_{b}R_{\mu\nu}{}^{ab}(J) \nonumber \\
     &=2v^{\mu}e_{a}^{\nu}\lambda_{b}e_{\rho}^{a}\ e^{\sigma\ b}\ \tilde{R}_{\mu\nu\sigma}{}^{\rho}
\end{align}
In order to preserve the invariance of the Ricci scalar under dipole transformations, we find the following constraint:
\begin{align}\label{constraint}
     \lambda_{b}\ v^{\mu}e_{a}^{\nu}e_{\rho}^{b}\ e^{\sigma\ a}\ \tilde{R}_{\mu\nu\sigma}{}^{\rho}=0
\end{align}
In 3 dimensions, the Riemann tensor can be written explicitly in terms of the Ricci tensor, Ricci scalar, and the metric:
\begin{equation}\label{riemann3d}
\tilde{R}_{\mu\nu\sigma\rho} = h_{\mu\sigma}\tilde{R}_{\nu\rho} - h_{\mu\rho}\tilde{R}_{\nu\sigma} - h_{\nu\sigma}\tilde{R}_{\mu\rho} + h_{\nu\rho}\tilde{R}_{\mu\sigma} - \frac{\tilde{R}}{2}(h_{\mu\sigma}h_{\nu\rho} - h_{\mu\rho}h_{\nu\sigma})
\end{equation}
where $R_{\mu\nu}$ is the Ricci tensor. Replacing \eqref{riemann3d}  into constraint \eqref{constraint}  we obtain the following constraint for Ricci tensor,
\begin{equation}
     v^{\mu} e^{\nu}_a  \tilde{R}_{\mu\nu} = 0,
\end{equation}
which will be satisfied  if the Ricci tensor is pure trace,  $\tilde{R}_{\mu\nu} = \frac{\tilde{R}}{3} h_{\mu\nu}$, according to the relation \eqref{kernel} and \eqref{vielbein}. It is worth noting that the constraint in Eq.~\eqref{constraint} represents a spacetime generalization of the spatial constraint reported in \cite{Slagle:2018kqf}. Furthermore, we emphasize that Eq.~\eqref{constraint} arises from the gauging algebra approach, specifically when a coupling term is included in the definition of covariant derivatives. In contrast, the constraint presented in \cite{Bidussi:2021nmp} is derived from a fracton gauge field theory formulated on a curved spatial background.

\subsection{Gauging the Center of Mass Algebra }\label{sec:COM}

In the previous subsection, we examined the fracton algebra that characterizes spacetime as Aristotelian spacetime. Now, we aim to identify a fundamentally different spacetime, which we will refer to as \textit{Center of Mass} spacetime, by extending the Aristotelian framework with the \textit{center of mass} operator as a new symmetry generator of spacetime. To achieve this goal, we include the center of mass operator $\hat{M_i}$ in the Aristotelian algebra as follows,
\begin{equation} \label{Fracton-alg 2}
  \begin{split} 
    \quad \{\hat{P_i},\hat{M_j}\}&=\delta_{ij}\hat{M} , \\
     \{\hat{P_i},\hat{J}_{jk}\} &=\delta_{ik}\hat{P_j}-\delta_{ij}\hat{P_k},\quad \{\hat{M_i},\hat{J}_{jk}\}=\delta_{ik}\hat{M_j}-\delta_{ij}\hat{M_k}, \\
     \{\hat{J}_{ij},\hat{J}_{kl}\} &=\delta_{il}\hat{J}_{jk}-\delta_{ik}\hat{J}_{jl}+\delta_{jk}\hat{J}_{il}-\delta_{jl}\hat{J}_{ik} .
     \end{split}
\end{equation}
The algebra \eqref{Fracton-alg 2} resembles the Carrollian algebra.  
Gauging this algebra is the same as gauging the Carrollian algebra, which begins with a connection $A_{\mu}$ as
\begin{align}\label{connection 2}
 A_{\mu}=\hat{M}\tau_{\mu}+\hat{P}_{a}e^{a}_{\mu}+\frac{1}{2}\hat{J}_{ab}\omega^{ab}_{\mu}+\hat{M}_{a}n^{a}_{\mu},
\end{align}
where the value of $\mu$ assumes $d+1$  distinct values, reflecting the fact that the system contains one time generator and $d$ space translation generators. Thus, the Riemann curvature tensor satisfies \cite{Hartong:2015xda}
\begin{align}
    R_{\mu\nu\sigma}^{ \ 
 \ \ \ \ \rho}=v^{\rho}e_{\sigma a}R_{\mu\nu}{}^{a}(M_{b})-e^{\rho}_{b}e_{\sigma \ a}R_{\mu\nu}{}^{ab}(J) \, ,
\end{align}
with an affine connection as:
\begin{align}
    \Gamma^{\sigma}_{\mu\nu}=-v^{\sigma}\partial_{\mu}\tau_{\nu}-\frac{1}{2}h^{\rho\sigma}(\partial_{\mu}h_{\nu\rho}+\partial_{\nu}h_{\mu\rho}-\partial_{\rho}h_{\mu\nu})+h^{\rho\sigma}(-\tau_{\nu}K_{\mu\rho}+\frac{1}{2}S_{\rho\mu\nu})-v^{\sigma}\Theta_{\mu\nu
    }.
\end{align}
One of the interesting aspects that can be observed in this case is that we can introduce the vector \( m^{\mu} \) to keep the vielbein \( \tau_{\mu} \) and \( e_{a}^{\ \mu} \) invariant under mass dipole transformations. 
\subsection{Fractons and Non-Lorentzian Gravities}
Fracton gravity is a theoretical framework for studying the behavior of particles known as fractons from a geometrical perspective. In the context of fracton gravity, we can describe the interaction between fractons using geometric concepts, analogous to how gravitational attraction is expressed in general relativity. A key aspect of fracton gravity is its connection to Mach's principle, which suggests that the inertia of a body is the result of its interaction with the rest of the universe's mass. In fracton gravity, an isolated particle is immobile, but it gains inertia from the presence of a large-scale distribution of other fractons, embodying Mach's principle at the effective level \cite{Pretko:2017fbf}. 

In the previous sections, we have established that fracton particles exhibit duality with non-Lorentzian particles. In this section, inspired by the duality of Fracton/Non-Lorentzian theories, we will explore the relationship between fracton gravity and the non-Lorentzian gravities.
There are two different limits of general relativity that can be defined at the level of Einstein-Hilbert action. The first limit is the ultrarelativistic limit, which leads to a Carroll gravity theory that is invariant under the Carroll symmetry. The second limit is a non-relativistic limit, the so-called Galilei limit, differing from the Newton-Cartan limit by the absence of a mass parameter and a central charge gauge field \cite{Bergshoeff:2017btm}.
The analysis of gauging the center of mass algebra \eqref{Fracton-alg 2} demonstrates that incorporating the conservation of dipole in the structure of spacetime results in a Carrollian spacetime. According to Carroll gravity, the Einstein-Hilbert action, is as follows \cite{Bergshoeff:2017btm}:
\begin{align}\label{Carroll gravity 1}
    S_{\textbf{Carroll/COM}}=\frac{1}{16\pi G_{C}}\int  e (2v^{\mu}e_{a}^{\nu}R_{\mu\nu}{}^{a}(M_{b})+e_{a}^{\mu}e_{b}^{\nu}R_{\mu\nu}{}^{ab}(J)),
\end{align}
where $R_{\mu\nu}{}^{a}(M_{b})$ is curvature 2-form of the center of mass/Carroll boost symmetry and $e=det(\tau_{\mu}, e_{\mu}^{a})$. In the second-order formalism it has the following form:
\begin{align}\label{Carroll gravity 2}
    S_{\textbf{Carroll/COM}}=\frac{1}{16\pi G_{C}}\int  e h^{\mu\nu}(R_{\mu\nu}+v^{\rho}\tau_{\sigma}R_{\mu\nu\rho}^{\sigma}),
\end{align}
The equations of motion will be obtained by varying $v^{\mu}$ and $h^{\mu\nu}$, 
\begin{equation}\label{eq:carroll-eom}
\begin{split}
    (\tau_{\lambda}h_{\mu}^{\sigma}-\frac{1}{2}\tau_{\mu}h_{\lambda}^{\sigma})h^{\nu\rho}R_{\sigma\nu\rho}^{\lambda}&=0,\\ R_{\mu\nu}-\frac{1}{2}h_{\mu\nu}\hat{R}&=0,
    \end{split}
\end{equation}
where $h_{\mu}^{\nu}=h_{\mu\rho}h^{\nu\rho}$ and $\hat{R}=h^{\mu\nu}(R_{\mu\nu}+v^{\rho}\tau_{\sigma}R_{\mu\nu\rho}^{\sigma})$.
An interpretation of equations \eqref{eq:carroll-eom} is a form of gravity that includes the conservation of the center of mass.
\section{Discussion and outlook}\label{conclusion}
The study of fractons, characterized by their constrained mobility and unique excitation spectra, has become a vibrant area of research in condensed matter and high-energy physics \cite{Pretko:2020cko, Nandkishore:2018sel, Gromov:2022cxa}. Initial investigations, pioneered by Chamon \cite{Chamon:2004lew}, focused on exactly solvable spin models, leveraging tools from quantum information theory to understand the exotic properties of these systems. These efforts aimed to elucidate the nature of quantum glassiness and explore the potential of fracton immobility for robust quantum memories \cite{Villari:2023myb}. Concurrently, Pretko's work on symmetric tensor gauge theories \cite{Pretko:2016kxt} established connections between fractons and other areas of physics, including elasticity theory and gravity \cite{Pretko:2017kvd, Pretko:2017fbf}. The coupling of fractons to spatial two-index symmetric tensor gauge fields ($A_{ij}$) hinted at a deep connection between dipole conservation, spin-2 fields, and potentially gravity itself. Further research has explored the internal dynamics of fracton systems, seeking links to emergent gravitational phenomena \cite{Afxonidis:2023pdq, Pretko:2017fbf}.

More recently, the exploration of non-Lorentzian physics has provided a new perspective on fractons. Non-relativistic limits of relativistic theories, such as Galilean and Carroll symmetries, offer alternative frameworks for describing physical phenomena. The Galilean limit, well-established in non-relativistic quantum mechanics and effective field theories \cite{levy1965nouvelle, le1973galilean, Taylor:2008tg, Nishida:2007pj, Son:2008ye, Goldberger:2008vg}, and the Carroll limit, arising when the speed of light approaches zero \cite{levy1965nouvelle, Duval:2014uoa, Bergshoeff:2022eog, Figueroa-OFarrill:2022nui, Figueroa-OFarrill:2022pus}, have revealed intriguing connections to fracton physics. Specifically, prior works \cite{Figueroa-OFarrill:2023vbj, Figueroa-OFarrill:2024ocf} have initiated the study of duality between Carroll/Galilean particles and Fractons.

Building upon these foundations, our work explores the relationship between fractons and non-Lorentzian particles, particularly focusing on the gauge field couplings of fractons and the conditions influencing their mobility. We have demonstrated a new set of dualities between fractons in gauge fields and non-Lorentzian particles in electromagnetic fields. By systematically employing a limiting procedure, as described in Sections \ref{Non-Relativistic point particle} and \ref{Non-Relativistic point particle in EM}, we have derived the non-Lorentzian particle dynamics and established their dual descriptions on the fractonic side (for details, see Section \ref{FNLduality}). These findings extend previously established dualities and offer a deeper understanding of the correspondence between fractons and non-Lorentzian physics, particularly in scenarios involving coupling to fracton gauge fields. These perspectives are further extended to include couplings between geometry and fracton gauge fields, using gauging algebra methods to investigate the resulting geometric implications and derive the corresponding constraints.
 
Our key findings and contributions include:

\begin{itemize}
    \item We developed a comprehensive framework for fracton dynamics coupled with gauge fields, addressing both scalar charge and vector charge theories. The corresponding evolution equations were derived, and the mobility characteristics of fractons within these systems were examined. 
    \item By constructing a systematic limiting approach, we analyzed free non-Lorentzian particles and successfully rederived Time-like Galilean and Space-like Carroll particles as emergent behaviors arising within these limiting regimes. This contrasts with previous studies, which introduced such particles via alternative methodologies. We, then, extended this framework to include couplings with gauge fields, classifying four distinct sectors of non-Lorentzian particle dynamics: Electric Charged Carroll Particles, Magnetic Charged Carroll Particles, Magnetic Charged Galilean Particles, and Electric Charged Galilean Particles. Among these, Electric Charged Carroll Particles had been previously discussed in \cite{Marsot:2022imf, Marsot:2021tvq}.
    \item We propose a new set of dualities between fractons and non-Lorentzian particles in gauge fields, as detailed in Subsection~\ref{FNLduality}.
    \item The symmetry principles governing fractons in gauge fields were extended to the algebraic level, constructing and introducing the Fracton-Maxwell and Vector Charge-Maxwell algebras (see Sections \ref{Fracton-Maxwell-algebra} and \ref{Vector-charge-algebra}), which share some structural similarities with Carroll and Galilei-Maxwell algebras.
    \item Using gauging algebra techniques, we explored geometric realization of fracton symmetry. This included the formulation of curvature invariants and analysis of curvatures under dipole symmetry, considering two distinct scenarios of gauge field coupling with background geometry. In cases where fractonic gauge fields interact with background geometry, the Aristotelian background is restricted by a constraint that ensures compatibility with dipole symmetry. This constraint (Eq.~\eqref{constraint}) is a spacetime generalization of those reported in \cite{Slagle:2018kqf} and \cite{Bidussi:2021nmp}, producing their results in three dimensions and imposes that the background Aristotelian geometry must be an Einstein manifold. 
    \item Additionally, we introduce an extension of the Aristotelian algebra incorporating a center-of-mass operator (Eq.~\eqref{Fracton-alg 2}) in analogy with coupling of fracton gauge field with background Aristotelian spacetime. We examine its spacetime curvature invariants and compare them with those derived from Carrollian geometry. These geometric frameworks offer an approach to understanding fracton phases and their potential connection to emergent gravitational phenomena, drawing inspiration from similar developments in non-Lorentzian geometries \cite{Bergshoeff:2022eog, Figueroa-OFarrill:2022nui}.
\end{itemize}

This work opens several avenues for future research. It would be interesting to explore the implications of our findings for specific condensed matter systems that exhibit fracton behavior and investigating dualities for higher-rank fracton phases (e.g., quadrupole-conserving systems) and their correspondence with non-Lorentzian theories. Furthermore, exploring the connections between fracton geometry and non-Lorentzian geometry could lead to a deeper understanding of emergent gravity and unconventional quantum field theories in condensed matter physics. The exploration of these connections promises to further illuminate the rich and complex landscape of fracton physics and its relation to the broader theoretical framework of non-Lorentzian physics.

\acknowledgments
We would like to thank M.M.~Sheikh-Jabbari, H.~Afshar, E. Have for their valuable comments and discussions. 
\appendix
\section{Fracton - Maxwell theory}
For scalar charge theory, the generalized fracton Maxwell-like equations can be expressed as follows \cite{Pretko:2016lgv}:
\begin{align}
    \partial_{i}\partial_{j}E^{ij}&=0,\\
    \partial_{i}B^{ij}&=0,\\
    \epsilon^{iab}\partial_{a}E^{j}_{b}-\partial_{t}B^{ij}&=0,\\
    \frac{1}{2}(\epsilon^{iab}\partial_{a}B^{j}_{b}+\epsilon^{jab}\partial_{a}B^{i}_{b}+\partial_{t}E^{ij})&=0.
\end{align}


\addcontentsline{toc}{section}{References}
\bibliographystyle{fullsort.bst}
\bibliography{biblio}

\end{document}